\documentclass[aps,twocolumn,secnumarabic,balancelastpage,amsmath,amssymb,nofootinbib, prl,10pt]{revtex4-2}

\usepackage{graphicx}
\usepackage{hyperref} 
\usepackage{color}
\usepackage{natbib}

\newcommand{\be}{\begin{equation}} 
\newcommand{\ee}{\end{equation}}
\newcommand{\bes}{\begin{equation*}}
\newcommand{\ees}{\end{equation*}}

  \usepackage[utf8]{inputenc}

\usepackage{aas_macros}

\newcommand{\refreply}[1]{\textcolor{black}{}}

\begin{document}
\title[Lyman-alpha forest and Wavelet scattering transform]{Forecast Cosmological Constraints with the 1D Wavelet Scattering Transform and the Lyman-alpha forest}
\author{
Hurum Maksora Tohfa$^{1,2}$}
\email{htohfa@uw.edu}
\author{Simeon Bird$^{1}$}
\email{sbird@ucr.edu}
\author{Ming-Feng Ho$^{1}$}
\author{Mahdi Qezlou$^{1}$}
\author{Martin Fernandez$^{1,3}$}
\affiliation{
\small{$^{1}$Department of Physics \& Astronomy, University of California Riverside, Riverside, CA 92521, United States}\\
\small{$^{2}$Department of Astronomy, University of Washington,  Seattle, WA 98195, United States} \\
\small{$^{3}$Department of Atmospheric Science, Colorado State University, Fort Collins, CO, United States} 
}
\date{\today}

\begin{abstract}
We make forecasts for the constraining power of the 1D Wavelet Scattering Transform (WST) when used with a Lyman-$\alpha$ forest cosmology survey. Using mock simulations and a Fisher matrix, we show that there is considerable cosmological information in the scattering transform coefficients not captured by the flux power spectrum. We estimate mock covariance matrices assuming uncorrelated Gaussian pixel noise for each quasar, at a level drawn from a simple lognormal model. The extra information comes from a smaller estimated covariance in the first-order wavelet power, and from second-order wavelet coefficients which probe non-Gaussian information in the forest. Forecast constraints on cosmological parameters from the WST are more than an order of magnitude tighter than for the power spectrum, shrinking a $4D$ parameter space by a factor of $10^6$. Should these improvements be realised with DESI, inflationary running would be constrained to test common inflationary models predicting $\alpha_s = - 6\times 10^{-4}$ and neutrino mass constraints would be improved enough for a $5-\sigma$ detection of the minimal neutrino mass.
\end{abstract}
\maketitle
\section{Introduction}

%what's lyman-alpha and why lyman-alpha
An important probe of small-scale structures is the Lyman-$\alpha$ forest, absorption features of neutral hydrogen. % atoms at rest-frame 1215.67\AA $ $. T
The Lyman-$\alpha$ forest allows us to explore fundamental questions about the Universe, such as the nature of dark matter, the total neutrino mass, and the physics of the inflationary era \cite{2005PhRvD..71j3515S,2015JCAP...11..011P,2017PhRvD..96b3522I,2017JCAP...06..047Y}, as well as measure the thermal history of the intergalactic medium (IGM) \cite{2002ApJ...578L...5T}. However, as the Lyman-$\alpha$ forest probes the non-linear regime, hydrodynamic simulations are needed to accurately predict its behavior under different cosmological parameters. %%experiments that do lyman-alpha
There are a variety of Lyman-$\alpha$ forest surveys \citep[e.g.~][]{2013A&A...559A..85P, 2017MNRAS.466.4332I, 2019JCAP...07..017C, 2022MNRAS.509.2842K}, with the largest statistical sample being from the extended Baryon Oscillation Spectroscopic Survey (eBOSS) \cite{2019MNRAS.489.2536D}. %accompanied by an average uncertainty of $6-18 \%$ for higher-resolution surveys.
The ongoing Dark Energy Spectroscopic Instrument (DESI)\cite{2016arXiv161100036D, 2023arXiv230606308D}, and WEAVE-QSO \cite{2016sf2a.conf..259P} will vastly increase the number of spectra. However, SDSS and DESI will be systematics dominated in the flux power spectrum, suggesting a need for new summary statistics, especially those which can extract non-Gaussian information from the non-linear density field.

%While the power spectrum is optimal for a Gaussian field, non-linear density fields are non-Gaussian. There may thus be substantial information within in the forest that can only be captured by summary statistics other than the power spectrum.

%why not just higher order summary statistics
Higher order n-point functions such as the bispectrum have been calculated for galaxy and weak lensing surveys \cite{2002A&A...389L..28B, 2003MNRAS.344..857T,2011MNRAS.410..143S, 2014MNRAS.441.2725F, 2023A&A...672L..10A, 2023MNRAS.522..606P}.
However, they suffer from increased variance and reduced robustness to outliers \cite{Welling2005RobustHO}. In highly non-Gaussian distributions with heavy tails, there can be significant fluctuations or noise in the data that make it difficult to extract meaningful features related to interactions/correlations. These fluctuations become more pronounced as we move toward higher orders and dilute important features \cite{2021PhDT........26C}.

%WST 
Here we investigate using the wavelet scattering transform for Lyman-$\alpha$ forest cosmology. The wavelet scattering transform (WST) was initially proposed by Mallat for signal processing \cite{2014ITSP...62.4114A}. In recent years, it has been used successfully in several signal processing fields \cite{Andn2011MultiscaleSF, 6619007}.
%The WST uses predefined filter banks that are similar to n-point statistics and so provide interpretability and transparency. 
Each order of the WST is fully deterministic, mathematically robust, has clearly interpretable structures, and does not require any training of the model \cite{2011arXiv1101.2286M, 2020MNRAS.499.5902C}. The WST convolves signals with a complex-valued wavelet at different scales, selecting Fourier modes around a central frequency and thus separating information from different scales.
% It then takes the modulus of the result of this convolution operation.
This first order WST is conceptually similar to the power spectrum, but may be applied repeatedly to extract higher order summary statistics.

%people have used wst before
The WST has been applied to weak lensing shear maps, where it has been shown to outperform the power spectrum and achieve similar constraining power to neural network techniques \cite{2021MNRAS.507.1012C}. 
While forecasts suggest that Convolutional Neural Networks (CNNs) can improve cosmological inference over two-point statistics, particularly for weak lensing data \cite{2019MNRAS.490.1843R}, other forecasts have suggested that scattering architectures are able to outperform a CNN \cite{2022mla..confE..40P}, and are more interpretable \cite{2012arXiv1203.1513B}. The WST has also been applied to mock 21-cm data, where it forecast tighter constraints than the 3D spherically averaged power spectrum \cite{2022MNRAS.513.1719G}. Finally, 3D WST has been shown to preserve 50\% more of the information in an idealised mock density field than the marked power spectrum \cite{2022PhRvD.105j3534V}.

%what we are doing
The wavelet power has been used to extract thermal history information from the Lyman-$\alpha$ forest by Ref.~\cite{2010ApJ...718..199L}, although constraints are similar to those from the flux power spectrum \cite{Gaikwad:2021}. We make forecasts for an extended wavelet-based analysis which both considers cosmological parameters and uses higher order WST coefficients to extract non-Gaussian information. We use the Fisher information matrix to make forecast constraints from a mock survey for both the WST and the power spectrum, finding that the WST substantially improves constraining power. Our analysis suggests a novel approach for extracting cosmological information from the Lyman-$\alpha$ forest.
%introducing WST

%In several of these areas, WST has reached comparable performance to CNN. WST is very similar to CNN as it involves the decomposition and convolution of signals into wavelets. However, unlike CNN, WST does not require the model to be trained. 

%outline Not for letter.
% The letter is structured as follows: in section \ref{2}, we discuss the Lyman-$\alpha$ forest simulations that we use. We give an overview of the WST and power spectrum in section \ref{3}, followed by the Fisher matrix formalism that we use for this work in section \ref{4}. We present our results and discuss them in section \ref{5} and draw our final conclusions in section \ref{6}.

\section{ Mock Data for analysis}
\label{2}

We use $10$ hydrodynamic simulations, described fully in Ref.~\cite{2019JCAP...02..050B}. Since this paper is aimed at making forecasts for the constraining power of different summary statistics rather than making a comparison to observational data, we use relatively fast, small simulations containing 2 $\times 256^3$ dark matter and Smoothed-Particle Hydrodynamics particles within a $40$ Mpc/h box. \refreply{The low resolution of these simulations affects the flux power spectrum at the $10-20\%$ level at $k = 0.02$ s/km \cite{2022MNRAS.517.3200F}. We pick parameters for a central reference simulation, and every other simulation varies exactly one parameter from this reference. For this pilot forecast, our simulations are not fully resolved in either resolution or box size. Ref.~\cite{2023arXiv230605471B} ran a suite of much larger simulations, but their use of a Latin Hypercube design makes computing the Fisher matrix gradients more complex.}

Fully hydrodynamic simulations were performed with the simulation code MP-Gadget \cite{yu_feng_2018_1451799}.
%The gravitational dynamics of the simulations are calculated using a Fourier transform on large scales and on small scales by walking a distributed Barnes-Hut tree structure \cite{1986Natur.324..446B}. We use the same transfer function for dark matter and baryonic particles and initialise at $z=99$.
We generate 32\,000 spectra Lyman-$\alpha$ absorption spectra from each simulation at $z=4, 3, 2$ with ``fake\_spectra'' \cite{2017ascl.soft10012B}. Spectra are generated parallel to the x-axis of the simulation box. We approximate a realistic survey by adding Gaussian noise and smoothing with a Gaussian filter with width corresponding to the spectral resolution of eBOSS. \refreply{We chose eBOSS as it is the lowest resolution large Lyman-$\alpha$~survey and thus we would expect it to have the least small-scale information, making our estimate of the amount by which constraints will improve conservative.}
%This smoothing helps to reproduce the instrumental effects that are present in real observations, such as the finite resolution of the instrument and any residual noise. We also add Gaussian noise to our simulations for generating a covariance matrix which will be further discussed in Section \ref{4}.

\begin{figure*}
  \centering
  \includegraphics[width=\textwidth]{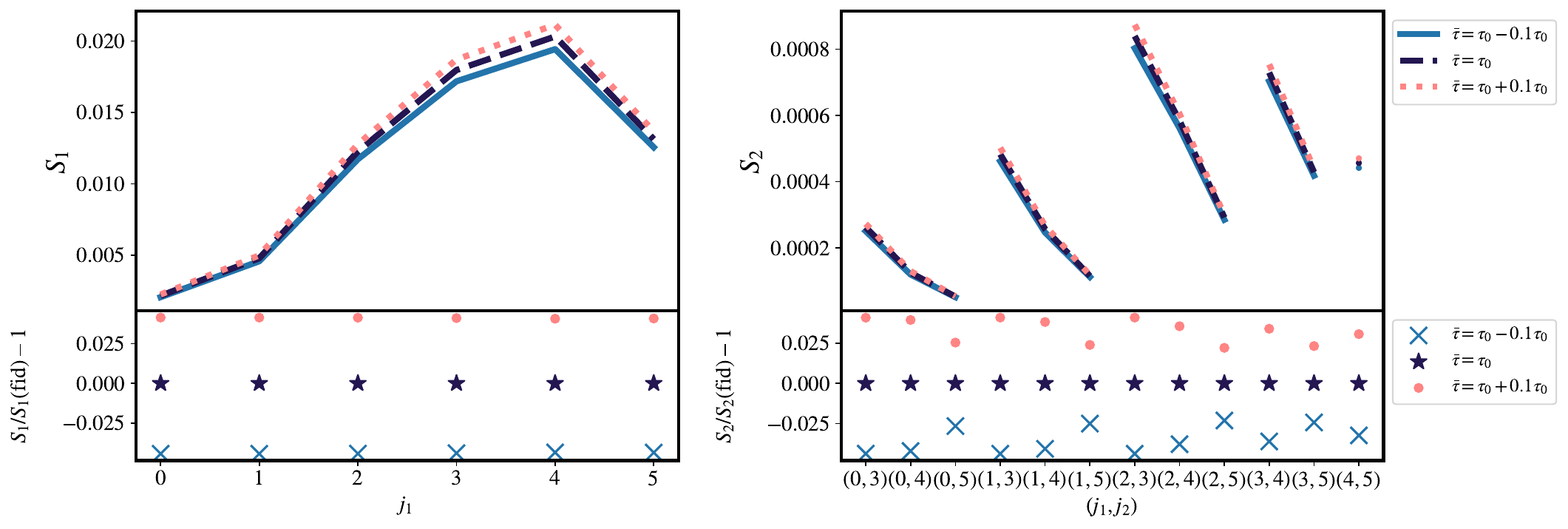}
  \caption{%\spb{We don't say anywhere what we mean by 'WST'. I think we mean the scattering coefficients $S_x$, averaged over $x$ and $y$, but this needs to be explicit.} 
  Dependence of the scattering coefficients described in Equation.~\ref{eq:wst} on changes in the mean flux at z$=2$. The top panels show the first and second order scattering coefficients and the bottom panels show the fractional change in mean flux with respect to the observed model.  We see about a $2.5\%$ change in scattering coefficients as the value of the mean flux changes by $10\%$.
  }
  \label{mean_flux}
\end{figure*}

Our simulation suite includes $4$ parameters: the spectral index, $n_P$, and amplitude, $A_P$ of the primordial power spectrum. The power spectrum is given by the equation:
\begin{equation}
P(k)=A_{\mathrm{P}}\left(\frac{k}{k_P}\right)^{n_P-1},
\end{equation}
where $k_P = 2\pi/8 = 0.78$ Mpc$^{-1}$.
%The amplitude of the power spectrum at $k=0.05$ h/Mpc, $A_s$ is related to $A_p$ via the relation:
%
%\begin{equation}
%A_s^{\mathrm{CMB}}=\left(\frac{1.6 \times 10^{-2}}{2 \pi}\right)^{n_P-1} %A_{\mathrm{P}}.
%\end{equation}
We include parameters to model uncertainty in the thermal history of the IGM \cite{2004MNRAS.354..684V}. We rescale the photo-heating rate by a density-dependent factor, $\tilde{\epsilon}=H_A \epsilon \Delta^{H_S}$. $\epsilon$ is the photo-heating rate, $\Delta$ is the overdensity of the IGM, $H_A$ controls the IGM temperature at mean density, and $H_S$ controls the slope of the temperature-density relation.

%In order to extract the most information from the Lyman-$\alpha$ forest, it is important to model it accurately and account for various sources of uncertainty. 
We add an extra parameter in post-processing for the observed mean flux in the forest, which is proportional to the overall ionization fraction of neutral hydrogen. We rescale our spectra to have the same mean flux by multiplying the optical depth in each spectral pixel by a constant factor. The mean optical depth follows the power law redshift evolution from Ref.~\cite{2007MNRAS.382.1657K}:
\begin{equation}
\tau = 2.3 \times 10^{-3}(1+z)^{3.65}.
\end{equation}

\section{The 1D Wavelet Scattering Transform}
\label{3}

We represent a spectrum by its input flux field $\mathcal{F}(x)$, a 1D array with velocity coordinate $x$. The 1D flux power spectrum is
\begin{align}
    P(k) &= \left\langle|\widehat{\mathcal{F}}(x)|^2\right\rangle = \left\langle\left|\mathcal{F}(x) \star \psi_k(x)\right|^2\right\rangle\,.
\end{align}
where $\psi_k = e^{-ikx}$ and $\star$ is a convolution. We average over all spectra in the box, denoted by $\langle \rangle$. \refreply{Thus both our summary statistics are sensitive only to line of sight information and integrate out transverse correlations.}

We define the WST coefficients via recursive convolution of the field $\mathcal{F}(x)$ with a series of wavelets with different numbers of octaves (denoted by Q) of different scales (j). Each wavelet in the wavelet family maintains an identical shape but varies in scale and orientation.
After convolution, we take the modulus of the generated fields and use low pass filters to smooth out the high-frequency components of the signal. We set  $Q = 1$ to employ a dilation factor of 2, making the wavelet's real-space size roughly equivalent to $2^J$ pixels \cite{2020MNRAS.499.5902C}, where $J$ denotes the largest physical scale possible. By averaging the resultant fields, we obtain scattering coefficients that describe the statistical characteristics of the input flux.
The zeroth, first, and second order scattering coefficients are defined as:
\begin{align}
\label{eq:wst}
S_0  &=  \langle \mathcal{F}(x) \star \phi[0] \rangle  \\ 
S_1\left(j_1 \right)  &=  \langle |\mathcal{F}(x) \star \psi^{j_1}| \star \phi[1] \rangle  \\ 
S_2\left(j_1, j_2 \right) &=   \langle || \mathcal{F}(x) \star \psi^{j_1}| \star \psi^{j_2}| \star \phi[2] \rangle
\end{align}
Here $\psi^{j}$ are Morlet wavelet filters and $\phi[n]$ are low-pass filters.
%$n$ denotes the order of the scattering transform.
We compute the coefficients for $0 \leq j \leq J $, take their modulus, and average over all sightlines, as denoted by $\langle \rangle$. Lower `$j$' values correspond to smaller scales, meaning they oscillate more slowly and capture more detailed small-scale structures. Higher-order scattering coefficients are computationally expensive \cite{2013arXiv1311.4104B}, so we limit ourselves to scattering coefficients up to the second order. We set $J = 5$, as we found that this extracted most of the information from our simulations, \refreply{although bigger boxes may need increased $J$}.
\begin{figure*}
  \centering
  \includegraphics[width=\textwidth]{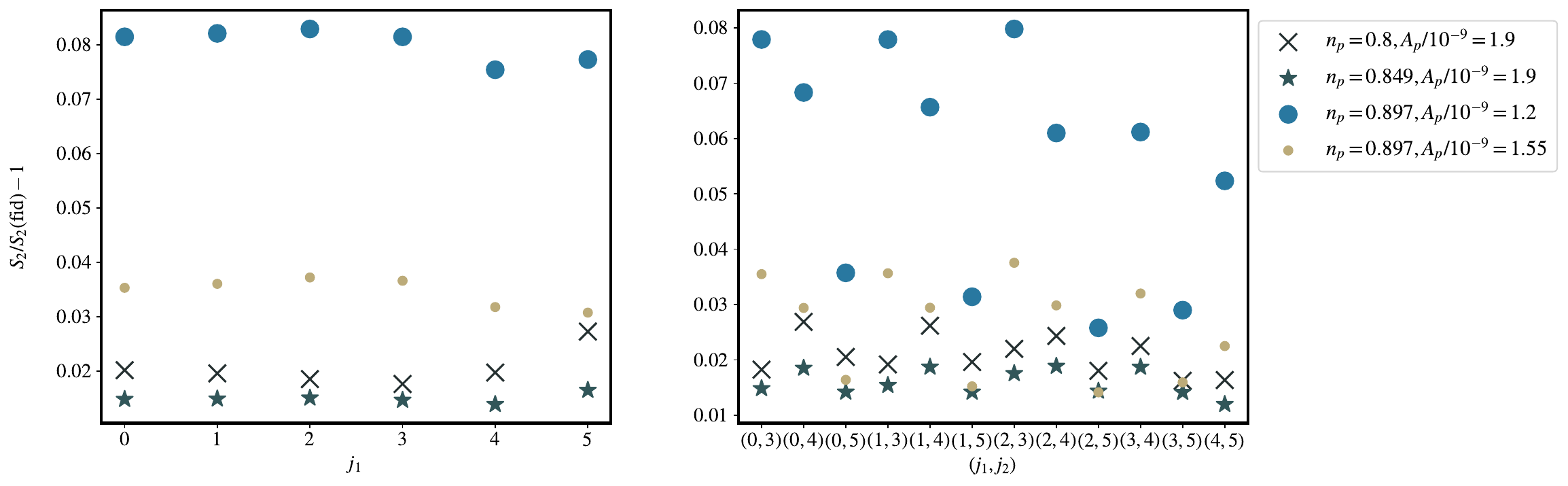}

  \caption{Change in the wavelet scattering coefficients defined in Equation~\ref{eq:wst} as we change our two cosmological parameters: $n_P$, $A_P$ at z$=2$. We show the fractional change in the cosmological parameters with respect to our fiducial model, similar to Fig. \protect\ref{mean_flux}.  Our fiducial cosmology is  defined as $n_P = 0.897$, $A_P =$ $1.9 \times 10^{-9} $, $H_S = -0.3$, $H_A = 0.9$. 
  }
  \label{cosmo_param}
\end{figure*}

The zeroth-order coefficient is acquired by applying a
low-pass filter, $\phi [J]$ to the input field,  $\mathcal{F}(x)$. It thus represents the local average of the field, which, for the Lyman-$\alpha$ spectra, corresponds to the local intensity of the transmitted flux. We expect this to be highly degenerate with the mean flux and continuum fitting and so to be conservative we focus on the first and second order coefficients.
%The resulting field is downsampled by a factor of $2^J$ prior to being transformed back into real space.

We use the publicly available KYMATIO package for generating 1D scattering coefficients \cite{2018arXiv181211214A}. \refreply{Conventionally, following the convolution, scattering coefficients are downsampled by $2^j$. After a low-pass filter, they are further downsampled by $2^{(J-j)}$ before inverse Fourier transforming back to real space, ensuring all coefficients are downsampled by $2^J$ and thus share the same coarsest resolution.
Second order coefficients are computed by applying a second convolution and modulus operation to the first order field (before low-pass filtering). A similar downsampling process to the first order coefficients is done, downsampling the new field by $2^{(j_1-j_2)}$ during convolution with the first-order field and further downsampling by $2^{(J - j_2)}$ using the low-pass filter.}

% The modulus converts fluctuations into their local  strengths. This generally has lower frequencies than the original fluctuations. Thus, modulus scatters information and energy from high frequencies and low frequencies \cite{2021PhDT........26C}.
The WST coefficients are Lipschitz continuous to deformation, meaning that similar fields differing by small deformations are also similar in terms of their scattering coefficient representation \cite{2020MNRAS.499.5902C}. This makes the scattering characterization stable even when there are slight variations or noise in the data.

%Higher-order coefficients can be derived by iteratively repeating the same process. However, the impact of larger scale second-order coefficients is relatively small. 

%By using linear operations like convolution followed by non-linear operations like taking moduli, WST can extract more meaningful information from the data while still preserving desirable properties such as stability under deformation. 

\section{Fisher Matrix Formalism}
\label{4}
We use the Fisher information approach to forecast the parameter constraints that we could achieve with a WST analysis. The Fisher matrix is defined as the second derivative of the log-likelihood function, $\ln \mathcal{L} ( p | d, M)$ around the maximum likelihood location, where $p$ is the 
parameter value of a model, and $d$ is the data.  Under this definition, we define the Fisher matrix as: 

\begin{equation}
F_{ij}=\left\langle\frac{\partial^2 \ln \mathcal{L}}{\partial p_i^2}\right\rangle= \frac{\partial \boldsymbol{S}}{\partial p_i} \cdot {\Sigma}^{-1} \cdot \frac{\partial \boldsymbol{S}}{\partial p_j}.
\end{equation}

Following Refs.~\citep{Lee12, Qezlou22}, we model the covariance $\boldsymbol{\Sigma}$ by adding Gaussian noise to the sightlines from our fiducial simulation. The continuum-to-noise ratio (CNR) for each simulated spectrum is sampled from a log-normal distribution, whose parameters are chosen to fit to the noise in the DR16Q observed spectral sample \footnote{\url{https://live-sdss4org-dr16.pantheonsite.io/algorithms/qso_catalog}}\citep{SDSS-DR16Q}:
\begin{equation}
    \mathrm{log(CNR)} \sim \mathcal{N}(\mu=0.53, 
\sigma=0.36)
\label{eq:cnr}
\end{equation}
Each spectrum, $i$ has a CNR value, CNR$_i$, drawn from the distribution in Eq.~\ref{eq:cnr}. Gaussian noise, $\epsilon_i$, realising CNR$_i$ is then added to each pixel:
\begin{align}
    \mathcal{F}(x)_j' &= \mathcal{F}(x)_j + \epsilon_i, \\
    \epsilon_i &\sim \mathcal{N}(\mu = 0, \sigma^2 = \mathrm{CNR}_i).
\end{align}
In this work, we do not consider the errors from incomplete QSO continuum fitting as they are subdominant to the CNR \citep{Lee12}. \refreply{However, higher order scattering coefficients might be sensitive to continuum subtraction.} A comprehensive treatment of continuum errors is deferred to future work.

%\mfh{Perhaps add a reference here explain why log-normal.}
We generate the covariance matrix $\Sigma$ by taking the variance of the scattering coefficients (or power spectrum), defined as: 
\begin{equation}
    \Sigma_{j k}=\frac{1}{N-1} \sum_{i=1}^N\left(X_{i j}-\bar{X}_j\right)\left(X_{i k}-\bar{X}_k\right)\,.
\end{equation} 
Here $X$ is the summary statistic, either the WST coefficients or the 1D flux power spectrum. We use $2000$ random noise realizations.

\section{Results}
\label{5}
In this Section, we describe the sensitivity of the WST and the 1D flux power spectrum to the parameters $A_P$, $n_P$, $H_S$ and $H_A$, as well as the mean flux $\tau_\mathrm{eff}$. The scattering transform with $J=5$ generates a total of $19$ scattering coefficients ($1$ zeroth order, $6$ first order, and $12$ second-order coefficients). We do not include the zeroth order coefficient in our analysis.
%The zeroth order coefficient is the local average of the field and so in real observational data would likely be dominated by continuum fitting.

Figure~\ref{mean_flux} shows the sensitivity of WST to changes in the mean flux.
%The mean flux is set by post-processing the spectra.
We computed the first and second order scattering transforms with a mean flux changed by $10\%$ in post-processing. The top panel shows the scattering coefficients and the bottom panel shows fractional change in the WST coefficients.
A 10$\%$ change in $\tau_\mathrm{eff}$ changes the first and second-order WST coefficients by $\sim 2.5\%$. For comparison, the 1D flux power spectrum is more sensitive to the mean flux, changing by about $7\%$.
 
\begin{figure*}[ht]
    \includegraphics[width=.9\textwidth]{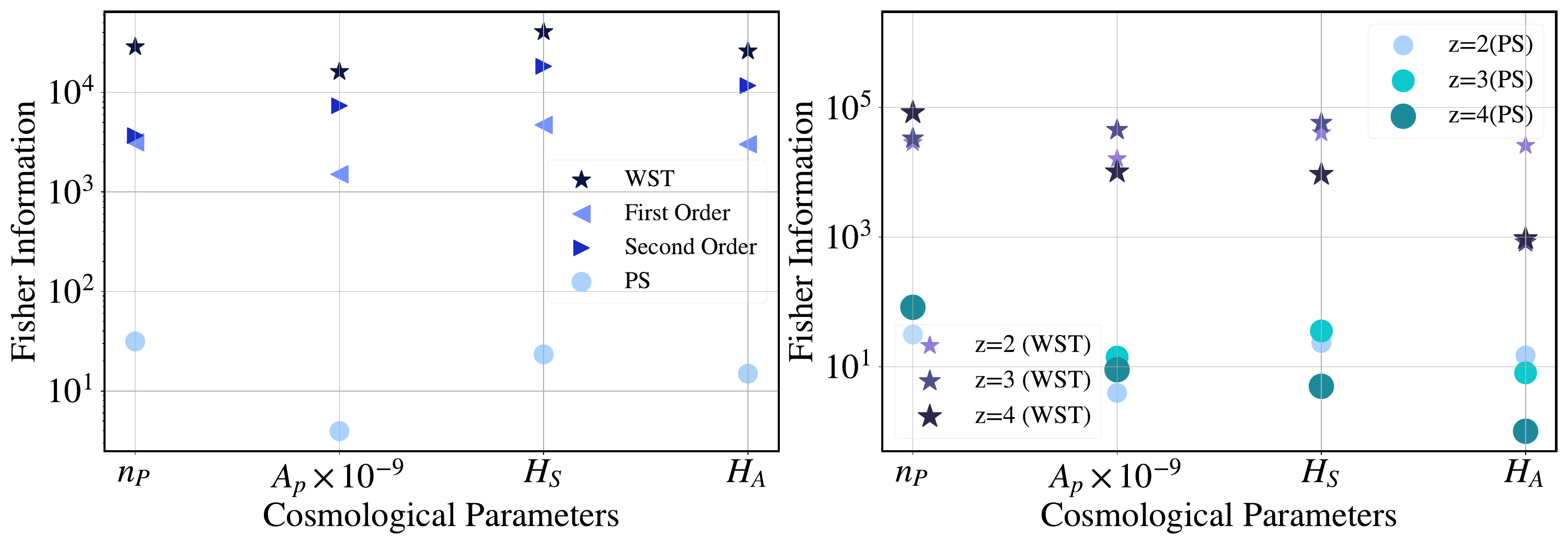}
    
    \caption{(Left) Fisher information from the power spectrum and wavelet scattering transform at $z=2$ for our four parameters. For the WST coefficients, we divide them into the contribution from first and second order coefficients, with second order coefficients generally being more constraining. (Right) The Fisher information for the combined WST coefficients and flux power spectrum for different redshift bins at $z=2,3,4$.
    }
    \label{cosmo param and redshift}
\end{figure*} 

Figure~\ref{cosmo_param} show how the WST coefficients are affected as we change cosmological
parameters. As in Figure~\ref{mean_flux}, the top panel shows the scattering coefficients and the bottom panel shows the fractional change in the WST coefficients with respect to the fiducial model. We see that changes in cosmological parameters indeed affect the scattering coefficients, signifying the presence of cosmological information.
Next, to compare the performance of WST with the power spectrum, we calculated the Fisher matrix.  Figure~\ref{cosmo param and redshift} shows the Fisher information in the WST coefficients and the power spectrum, for our four parameters $n_P$, $A_P$, $H_A$, $H_S$. The second-order fields contribute about 50\% extra information to the Fisher matrix over the first order fields. However, even the first order coefficients outperform the flux power spectrum. %, as they contain information on both amplitude and phase interactions.

%Even though the power spectrum and first-order WST coefficients are structurally similar, the first-order coefficients outperform the power spectrum. This may be because the power spectrum uses Fourier modes which are delocalized in real space and therefore lose spatial information. WST on the other hand uses localized kernels and captures information about both frequency and location. We find that $|\Sigma|$ for the power spectrum is larger than the scattering transform, implying a lower sensitivity to noise. Since the wavelet covers the whole Fourier space, the mean of the generated first order field, $I_1$, characterizes the average amplitude of Fourier modes but does not provide information about how these modes are spatially distributed or interact with each other. %\spb{This paragraph is about how the first order coefficients compare to the flux power spectrum. Let's split the result that there is more information in the second order coefficients into a second paragraph so it is clearer.} 

We can invert the Fisher matrix to estimate posterior uncertainties. For a parameter $X$, $\sigma_X = 1/ \sqrt{F_{ij}}$. Table~\ref{tab:uncertainty} shows $\sigma_X$ for the power spectrum and combined WST coefficients. The WST improves $\sigma_X$ by a factor of $30-60$.
%We find that for the power spectrum slope $n_P$,  the power spectrum can constrain $n_P$ with an error of $\sigma_{n_P} = 0.178$. In contrast, WST yields an improved constraint of $\sigma_{n_P} = 0.0059$.
\begin{table}[h]
\centering
\begin{tabular}{ c c c }
\hline 
Parameter & Power Spectrum & WST \\ 
\hline
$\Delta n_P$ & $0.178$ & $0.0059$ \\
$\Delta A_P / 10^{-9}$ & $0.503$ & $0.0079 $ \\
$\Delta H_S$ & $0.207$ & $0.0049$ \\
$\Delta H_A$ & $0.258$ & $0.0062$ \\
\hline
\end{tabular}
\caption{Estimated posterior uncertainties in parameters constraints obtained from both the power spectrum and the combined WST coefficients.\label{tab:uncertainty}}
\end{table}

These constraints are generated using a Fisher matrix rather than a full likelihood calculation. \refreply{Ref.~\cite{2023arXiv230903943F} analysed the eBOSS Lyman-$\alpha$ forest 1D flux power spectrum and found $\Delta n_P \approx 0.013$, $\Delta A_P /10^{-9} = 0.12$, indicating generally tighter constraints than our Fisher matrix prediction. However, that analysis includes larger scales absent in our small test boxes, which are particularly useful for constraining the power spectrum slope.} On the other hand, our Fisher matrix estimates neglect parameter degeneracies. The pivot scale $k_P$ is chosen to minimize correlation between $A_P$ and $n_P$, but there is some correlation between the cosmological and thermal parameters, visible in Figure 8 of Ref.~\cite{2023arXiv230903943F}.

Figure \ref{cosmo param and redshift} also shows the redshift dependence of both summary statistics. The constraining power of both the WST coefficients and the flux power spectrum varies similarly with redshift. We find that redshift $3$ contains the most information in both WST and the flux power spectrum for all parameters except $H_A$, which controls the mean IGM temperature and is most constrained by $z=2$.

\section{Conclusion}
\label{6}

\refreply{We have examined the power of the wavelet scattering transform (WST) to constrain cosmological parameters using the Lyman-$\alpha$ forest. Using small pilot simulations and a Fisher analysis, we show that the WST improves parameter constraints substantially. Our forecasts suggest an improvement by a factor of $10^6$ in the posterior volume of the 4D parameter space, an average of $42$ times greater constraints in each parameter.}

\refreply{These improved constraints come from two effects. First, the first-order WST is less sensitive to the mean flux, and more local in real space than the Fourier basis of the flux power spectrum. Second, constraints are improved by non-Gaussian information from the second-order WST.
These improved constraints would translate into tighter constraints on the scientific goals of the survey. Constraints on inflationary running, and thus also axion dark matter \cite{2023arXiv231116377R} or early dark energy models \cite{2023PhRvL.131t1001G}, which are proportional to the uncertainty on $n_P$, would tighten. Tightening the current eBOSS constraints ($\Delta \alpha_s = 0.0022$ \cite{2023arXiv231116377R}) by a factor of $30$ would constrain $\Delta \alpha_s \sim 7\times 10^{-5}$. Importantly this would be sufficient to test the large number of currently successful inflationary models, associated with spontaneous symmetry breaking, that predict $\alpha_2 = - 6 \times 10^{-4}$ \cite{Planck2018Inflation}. The neutrino mass constraints forecast from DESI are $\Delta M_\nu = 0.039$ eV \cite{2014JCAP...05..023F} and an order of magnitude tighter constraints would imply a five sigma detection of the neutrino mass, although realising these bounds would depend on a successful extension of this work to the transverse correlations. }

Our Fisher matrix formalism does not account for degeneracies between the cosmological parameters. Ref.~\cite{2023arXiv230903943F} discusses the parameter degeneracies for the 1D flux power spectrum, in a model closely related to ours. The largest correlation was between the mean optical depth $\tau_\mathrm{eff}$ and the primordial power spectrum amplitude $A_P$. The WST is less sensitive to $\tau_\mathrm{eff}$ and so it is likely this correlation will be reduced. Most other correlations were relatively small, with a correlation coefficient  $< 0.4$. Our purpose here is to compare the WST as a summary statistic to the flux power spectrum; it is likely that the parameter correlations will be similar between the two statistics, and so our main conclusion that the 1D WST provides information not captured by the 1D flux power spectrum should be robust. Our small simulations also do not include the largest scales accessible to the survey. This particularly impacts our forecasts for $n_P$, which were substantially too pessimistic for the flux power spectrum, and are likely also too pessimistic for the WST.

\refreply{Our analysis included uncertainty from noise. However, we did not model
continuum fitting error. The reduced sensitivity to $\tau_\mathrm{eff}$ in WST suggests this may be less important than in the flux power spectrum, but it may still be significant.
In addition, systematic error from uncertainty in the spectrograph resolution is important in eBOSS \cite{2019JCAP...07..017C}, although less so for DESI \cite{2023MNRAS.526.5118R}.} There is no a priori reason to suppose that these and other systematics should affect the WST more than the flux power spectrum, but a robust analysis of systematic errors will be needed when using the WST to obtain parameter constraints from DESI. %\spb{Can you reword this paragraph to explain more carefully that we have neglected several possible systematic errors and will look at them closely in future work?}
%WST can help future cosmological surveys deal with this problem as we have shown WST's utility and stability in generating summary statistics at different redshifts, and in handling small-scale fluctuations.
There are many avenues for future work. A maximum likelihood estimator for the WST coefficients from noisy and sparse data will need to be constructed. Our Fisher matrix forecast is simplistic and we will need to build a full likelihood function using some form of emulator of simulation based inference. However, despite these limitations, our work opens up the possibility to improve parameter constraints from the Lyman-$\alpha$ forest by multiple orders of magnitude.

%These findings carry profound implications for future cosmological studies. As we continue to refine our understanding of the Universe, we require increasingly nuanced methods for analyzing data.
%The observed performance of WST in this study shows that WST will be an effective tool for future research, including experiments such as the Dark Energy Spectroscopic Instrument (DESI) and the WEAVE-QSO survey. Since these experiments will generate an unprecedented volume of high-quality data, it is essential to refine our tools and methods to take full advantage of these ambitious experiments. The results of our research provide a promising step in this direction.

%However, to constrain cosmological parameters from these surveys, a variety of future work is necessary. Most importantly, we will need to construct a maximum likelihood estimator for the WST coefficients from noisy and sparse data. In addition, our Fisher matrix forecast is simplistic and we will in future work investigate the WST more thoroughly, using an emulator and Markov Chain Monte Carlo analysis.

\begin{acknowledgments}

SB was supported by NASA-80NSSC21K1840. 
MQ was supported by NSF grant AST-2107821 and the UCOP Dissertation Year Program. MFH is supported by a NASA FINESST grant No. ASTRO20-0022.
Computing resources were provided by Frontera LRAC AST21005.
The authors acknowledge the Frontera computing project at the Texas Advanced Computing Center (TACC) for providing HPC and storage resources that have contributed to the research results reported within this paper.
Frontera is made possible by National Science Foundation award OAC-1818253.
Analysis computations were performed using the resources of the UCR HPCC, which were funded by grants from NSF (MRI-2215705, MRI-1429826) and NIH (1S10OD016290-01A1).

\end{acknowledgments}

\bibliographystyle{apsrev4-2}
\bibliography{main}

\begin{thebibliography}{45}%
\makeatletter
\providecommand \@ifxundefined [1]{%
 \@ifx{#1\undefined}
}%
\providecommand \@ifnum [1]{%
 \ifnum #1\expandafter \@firstoftwo
 \else \expandafter \@secondoftwo
 \fi
}%
\providecommand \@ifx [1]{%
 \ifx #1\expandafter \@firstoftwo
 \else \expandafter \@secondoftwo
 \fi
}%
\providecommand \natexlab [1]{#1}%
\providecommand \enquote  [1]{``#1''}%
\providecommand \bibnamefont  [1]{#1}%
\providecommand \bibfnamefont [1]{#1}%
\providecommand \citenamefont [1]{#1}%
\providecommand \href@noop [0]{\@secondoftwo}%
\providecommand \href [0]{\begingroup \@sanitize@url \@href}%
\providecommand \@href[1]{\@@startlink{#1}\@@href}%
\providecommand \@@href[1]{\endgroup#1\@@endlink}%
\providecommand \@sanitize@url [0]{\catcode `\\12\catcode `\$12\catcode `\&12\catcode `\#12\catcode `\^12\catcode `\_12\catcode `\%12\relax}%
\providecommand \@@startlink[1]{}%
\providecommand \@@endlink[0]{}%
\providecommand \url  [0]{\begingroup\@sanitize@url \@url }%
\providecommand \@url [1]{\endgroup\@href {#1}{\urlprefix }}%
\providecommand \urlprefix  [0]{URL }%
\providecommand \Eprint [0]{\href }%
\providecommand \doibase [0]{https://doi.org/}%
\providecommand \selectlanguage [0]{\@gobble}%
\providecommand \bibinfo  [0]{\@secondoftwo}%
\providecommand \bibfield  [0]{\@secondoftwo}%
\providecommand \translation [1]{[#1]}%
\providecommand \BibitemOpen [0]{}%
\providecommand \bibitemStop [0]{}%
\providecommand \bibitemNoStop [0]{.\EOS\space}%
\providecommand \EOS [0]{\spacefactor3000\relax}%
\providecommand \BibitemShut  [1]{\csname bibitem#1\endcsname}%
\let\auto@bib@innerbib\@empty
%</preamble>
\bibitem [{\citenamefont {{Seljak}}\ \emph {et~al.}(2005)\citenamefont {{Seljak}}, \citenamefont {{Makarov}}, \citenamefont {{McDonald}}, \citenamefont {{Anderson}}, \citenamefont {{Bahcall}}, \citenamefont {{Brinkmann}}, \citenamefont {{Burles}}, \citenamefont {{Cen}}, \citenamefont {{Doi}}, \citenamefont {{Gunn}}, \citenamefont {{Ivezi{\'c}}}, \citenamefont {{Kent}}, \citenamefont {{Loveday}}, \citenamefont {{Lupton}}, \citenamefont {{Munn}}, \citenamefont {{Nichol}}, \citenamefont {{Ostriker}}, \citenamefont {{Schlegel}}, \citenamefont {{Schneider}}, \citenamefont {{Tegmark}}, \citenamefont {{Berk}}, \citenamefont {{Weinberg}},\ and\ \citenamefont {{York}}}]{2005PhRvD..71j3515S}%
  \BibitemOpen
  \bibfield  {author} {\bibinfo {author} {\bibfnamefont {U.}~\bibnamefont {{Seljak}}}, \bibinfo {author} {\bibfnamefont {A.}~\bibnamefont {{Makarov}}}, \bibinfo {author} {\bibfnamefont {P.}~\bibnamefont {{McDonald}}}, \bibinfo {author} {\bibfnamefont {S.~F.}\ \bibnamefont {{Anderson}}}, \bibinfo {author} {\bibfnamefont {N.~A.}\ \bibnamefont {{Bahcall}}}, \bibinfo {author} {\bibfnamefont {J.}~\bibnamefont {{Brinkmann}}}, \bibinfo {author} {\bibfnamefont {S.}~\bibnamefont {{Burles}}}, \bibinfo {author} {\bibfnamefont {R.}~\bibnamefont {{Cen}}}, \bibinfo {author} {\bibfnamefont {M.}~\bibnamefont {{Doi}}}, \bibinfo {author} {\bibfnamefont {J.~E.}\ \bibnamefont {{Gunn}}}, \bibinfo {author} {\bibfnamefont {{\v{Z}}.}~\bibnamefont {{Ivezi{\'c}}}}, \bibinfo {author} {\bibfnamefont {S.}~\bibnamefont {{Kent}}}, \bibinfo {author} {\bibfnamefont {J.}~\bibnamefont {{Loveday}}}, \bibinfo {author} {\bibfnamefont {R.~H.}\ \bibnamefont {{Lupton}}}, \bibinfo {author} {\bibfnamefont {J.~A.}\ \bibnamefont {{Munn}}}, \bibinfo
  {author} {\bibfnamefont {R.~C.}\ \bibnamefont {{Nichol}}}, \bibinfo {author} {\bibfnamefont {J.~P.}\ \bibnamefont {{Ostriker}}}, \bibinfo {author} {\bibfnamefont {D.~J.}\ \bibnamefont {{Schlegel}}}, \bibinfo {author} {\bibfnamefont {D.~P.}\ \bibnamefont {{Schneider}}}, \bibinfo {author} {\bibfnamefont {M.}~\bibnamefont {{Tegmark}}}, \bibinfo {author} {\bibfnamefont {D.~E.}\ \bibnamefont {{Berk}}}, \bibinfo {author} {\bibfnamefont {D.~H.}\ \bibnamefont {{Weinberg}}},\ and\ \bibinfo {author} {\bibfnamefont {D.~G.}\ \bibnamefont {{York}}},\ }\href {https://doi.org/10.1103/PhysRevD.71.103515} {\bibfield  {journal} {\bibinfo  {journal} {\prd}\ }\textbf {\bibinfo {volume} {71}},\ \bibinfo {eid} {103515} (\bibinfo {year} {2005})},\ \Eprint {https://arxiv.org/abs/astro-ph/0407372} {arXiv:astro-ph/0407372 [astro-ph]} \BibitemShut {NoStop}%
\bibitem [{\citenamefont {{Palanque-Delabrouille}}\ \emph {et~al.}(2015)\citenamefont {{Palanque-Delabrouille}}, \citenamefont {{Y{\`e}che}}, \citenamefont {{Baur}}, \citenamefont {{Magneville}}, \citenamefont {{Rossi}}, \citenamefont {{Lesgourgues}}, \citenamefont {{Borde}}, \citenamefont {{Burtin}}, \citenamefont {{LeGoff}}, \citenamefont {{Rich}}, \citenamefont {{Viel}},\ and\ \citenamefont {{Weinberg}}}]{2015JCAP...11..011P}%
  \BibitemOpen
  \bibfield  {author} {\bibinfo {author} {\bibfnamefont {N.}~\bibnamefont {{Palanque-Delabrouille}}}, \bibinfo {author} {\bibfnamefont {C.}~\bibnamefont {{Y{\`e}che}}}, \bibinfo {author} {\bibfnamefont {J.}~\bibnamefont {{Baur}}}, \bibinfo {author} {\bibfnamefont {C.}~\bibnamefont {{Magneville}}}, \bibinfo {author} {\bibfnamefont {G.}~\bibnamefont {{Rossi}}}, \bibinfo {author} {\bibfnamefont {J.}~\bibnamefont {{Lesgourgues}}}, \bibinfo {author} {\bibfnamefont {A.}~\bibnamefont {{Borde}}}, \bibinfo {author} {\bibfnamefont {E.}~\bibnamefont {{Burtin}}}, \bibinfo {author} {\bibfnamefont {J.-M.}\ \bibnamefont {{LeGoff}}}, \bibinfo {author} {\bibfnamefont {J.}~\bibnamefont {{Rich}}}, \bibinfo {author} {\bibfnamefont {M.}~\bibnamefont {{Viel}}},\ and\ \bibinfo {author} {\bibfnamefont {D.}~\bibnamefont {{Weinberg}}},\ }\href {https://doi.org/10.1088/1475-7516/2015/11/011} {\bibfield  {journal} {\bibinfo  {journal} {\jcap}\ }\textbf {\bibinfo {volume} {2015}},\ \bibinfo {pages} {011} (\bibinfo {year} {2015})},\
  \Eprint {https://arxiv.org/abs/1506.05976} {arXiv:1506.05976 [astro-ph.CO]} \BibitemShut {NoStop}%
\bibitem [{\citenamefont {{Ir{\v{s}}i{\v{c}}}}\ \emph {et~al.}(2017{\natexlab{a}})\citenamefont {{Ir{\v{s}}i{\v{c}}}}, \citenamefont {{Viel}}, \citenamefont {{Haehnelt}}, \citenamefont {{Bolton}}, \citenamefont {{Cristiani}}, \citenamefont {{Becker}}, \citenamefont {{D'Odorico}}, \citenamefont {{Cupani}}, \citenamefont {{Kim}}, \citenamefont {{Berg}}, \citenamefont {{L{\'o}pez}}, \citenamefont {{Ellison}}, \citenamefont {{Christensen}}, \citenamefont {{Denney}},\ and\ \citenamefont {{Worseck}}}]{2017PhRvD..96b3522I}%
  \BibitemOpen
  \bibfield  {author} {\bibinfo {author} {\bibfnamefont {V.}~\bibnamefont {{Ir{\v{s}}i{\v{c}}}}}, \bibinfo {author} {\bibfnamefont {M.}~\bibnamefont {{Viel}}}, \bibinfo {author} {\bibfnamefont {M.~G.}\ \bibnamefont {{Haehnelt}}}, \bibinfo {author} {\bibfnamefont {J.~S.}\ \bibnamefont {{Bolton}}}, \bibinfo {author} {\bibfnamefont {S.}~\bibnamefont {{Cristiani}}}, \bibinfo {author} {\bibfnamefont {G.~D.}\ \bibnamefont {{Becker}}}, \bibinfo {author} {\bibfnamefont {V.}~\bibnamefont {{D'Odorico}}}, \bibinfo {author} {\bibfnamefont {G.}~\bibnamefont {{Cupani}}}, \bibinfo {author} {\bibfnamefont {T.-S.}\ \bibnamefont {{Kim}}}, \bibinfo {author} {\bibfnamefont {T.~A.~M.}\ \bibnamefont {{Berg}}}, \bibinfo {author} {\bibfnamefont {S.}~\bibnamefont {{L{\'o}pez}}}, \bibinfo {author} {\bibfnamefont {S.}~\bibnamefont {{Ellison}}}, \bibinfo {author} {\bibfnamefont {L.}~\bibnamefont {{Christensen}}}, \bibinfo {author} {\bibfnamefont {K.~D.}\ \bibnamefont {{Denney}}},\ and\ \bibinfo {author} {\bibfnamefont {G.}~\bibnamefont
  {{Worseck}}},\ }\href {https://doi.org/10.1103/PhysRevD.96.023522} {\bibfield  {journal} {\bibinfo  {journal} {\prd}\ }\textbf {\bibinfo {volume} {96}},\ \bibinfo {eid} {023522} (\bibinfo {year} {2017}{\natexlab{a}})},\ \Eprint {https://arxiv.org/abs/1702.01764} {arXiv:1702.01764 [astro-ph.CO]} \BibitemShut {NoStop}%
\bibitem [{\citenamefont {{Y{\`e}che}}\ \emph {et~al.}(2017)\citenamefont {{Y{\`e}che}}, \citenamefont {{Palanque-Delabrouille}}, \citenamefont {{Baur}},\ and\ \citenamefont {{du Mas des Bourboux}}}]{2017JCAP...06..047Y}%
  \BibitemOpen
  \bibfield  {author} {\bibinfo {author} {\bibfnamefont {C.}~\bibnamefont {{Y{\`e}che}}}, \bibinfo {author} {\bibfnamefont {N.}~\bibnamefont {{Palanque-Delabrouille}}}, \bibinfo {author} {\bibfnamefont {J.}~\bibnamefont {{Baur}}},\ and\ \bibinfo {author} {\bibfnamefont {H.}~\bibnamefont {{du Mas des Bourboux}}},\ }\href {https://doi.org/10.1088/1475-7516/2017/06/047} {\bibfield  {journal} {\bibinfo  {journal} {\jcap}\ }\textbf {\bibinfo {volume} {2017}},\ \bibinfo {eid} {047} (\bibinfo {year} {2017})},\ \Eprint {https://arxiv.org/abs/1702.03314} {arXiv:1702.03314 [astro-ph.CO]} \BibitemShut {NoStop}%
\bibitem [{\citenamefont {{Theuns}}\ \emph {et~al.}(2002)\citenamefont {{Theuns}}, \citenamefont {{Viel}}, \citenamefont {{Kay}}, \citenamefont {{Schaye}}, \citenamefont {{Carswell}},\ and\ \citenamefont {{Tzanavaris}}}]{2002ApJ...578L...5T}%
  \BibitemOpen
  \bibfield  {author} {\bibinfo {author} {\bibfnamefont {T.}~\bibnamefont {{Theuns}}}, \bibinfo {author} {\bibfnamefont {M.}~\bibnamefont {{Viel}}}, \bibinfo {author} {\bibfnamefont {S.}~\bibnamefont {{Kay}}}, \bibinfo {author} {\bibfnamefont {J.}~\bibnamefont {{Schaye}}}, \bibinfo {author} {\bibfnamefont {R.~F.}\ \bibnamefont {{Carswell}}},\ and\ \bibinfo {author} {\bibfnamefont {P.}~\bibnamefont {{Tzanavaris}}},\ }\href {https://doi.org/10.1086/344521} {\bibfield  {journal} {\bibinfo  {journal} {\apjl}\ }\textbf {\bibinfo {volume} {578}},\ \bibinfo {pages} {L5} (\bibinfo {year} {2002})},\ \Eprint {https://arxiv.org/abs/astro-ph/0208418} {arXiv:astro-ph/0208418 [astro-ph]} \BibitemShut {NoStop}%
\bibitem [{\citenamefont {{Palanque-Delabrouille}}\ \emph {et~al.}(2013)\citenamefont {{Palanque-Delabrouille}}, \citenamefont {{Y{\`e}che}}, \citenamefont {{Borde}}, \citenamefont {{Le Goff}}, \citenamefont {{Rossi}}, \citenamefont {{Viel}}, \citenamefont {{Aubourg}}, \citenamefont {{Bailey}}, \citenamefont {{Bautista}}, \citenamefont {{Blomqvist}}, \citenamefont {{Bolton}}, \citenamefont {{Bolton}}, \citenamefont {{Busca}}, \citenamefont {{Carithers}}, \citenamefont {{Croft}}, \citenamefont {{Dawson}}, \citenamefont {{Delubac}}, \citenamefont {{Font-Ribera}}, \citenamefont {{Ho}}, \citenamefont {{Kirkby}}, \citenamefont {{Lee}}, \citenamefont {{Margala}}, \citenamefont {{Miralda-Escud{\'e}}}, \citenamefont {{Muna}}, \citenamefont {{Myers}}, \citenamefont {{Noterdaeme}}, \citenamefont {{P{\^a}ris}}, \citenamefont {{Petitjean}}, \citenamefont {{Pieri}}, \citenamefont {{Rich}}, \citenamefont {{Rollinde}}, \citenamefont {{Ross}}, \citenamefont {{Schlegel}}, \citenamefont {{Schneider}}, \citenamefont
  {{Slosar}},\ and\ \citenamefont {{Weinberg}}}]{2013A&A...559A..85P}%
  \BibitemOpen
  \bibfield  {author} {\bibinfo {author} {\bibfnamefont {N.}~\bibnamefont {{Palanque-Delabrouille}}}, \bibinfo {author} {\bibfnamefont {C.}~\bibnamefont {{Y{\`e}che}}}, \bibinfo {author} {\bibfnamefont {A.}~\bibnamefont {{Borde}}}, \bibinfo {author} {\bibfnamefont {J.-M.}\ \bibnamefont {{Le Goff}}}, \bibinfo {author} {\bibfnamefont {G.}~\bibnamefont {{Rossi}}}, \bibinfo {author} {\bibfnamefont {M.}~\bibnamefont {{Viel}}}, \bibinfo {author} {\bibfnamefont {{\'E}.}~\bibnamefont {{Aubourg}}}, \bibinfo {author} {\bibfnamefont {S.}~\bibnamefont {{Bailey}}}, \bibinfo {author} {\bibfnamefont {J.}~\bibnamefont {{Bautista}}}, \bibinfo {author} {\bibfnamefont {M.}~\bibnamefont {{Blomqvist}}}, \bibinfo {author} {\bibfnamefont {A.}~\bibnamefont {{Bolton}}}, \bibinfo {author} {\bibfnamefont {J.~S.}\ \bibnamefont {{Bolton}}}, \bibinfo {author} {\bibfnamefont {N.~G.}\ \bibnamefont {{Busca}}}, \bibinfo {author} {\bibfnamefont {B.}~\bibnamefont {{Carithers}}}, \bibinfo {author} {\bibfnamefont {R.~A.~C.}\ \bibnamefont
  {{Croft}}}, \bibinfo {author} {\bibfnamefont {K.~S.}\ \bibnamefont {{Dawson}}}, \bibinfo {author} {\bibfnamefont {T.}~\bibnamefont {{Delubac}}}, \bibinfo {author} {\bibfnamefont {A.}~\bibnamefont {{Font-Ribera}}}, \bibinfo {author} {\bibfnamefont {S.}~\bibnamefont {{Ho}}}, \bibinfo {author} {\bibfnamefont {D.}~\bibnamefont {{Kirkby}}}, \bibinfo {author} {\bibfnamefont {K.-G.}\ \bibnamefont {{Lee}}}, \bibinfo {author} {\bibfnamefont {D.}~\bibnamefont {{Margala}}}, \bibinfo {author} {\bibfnamefont {J.}~\bibnamefont {{Miralda-Escud{\'e}}}}, \bibinfo {author} {\bibfnamefont {D.}~\bibnamefont {{Muna}}}, \bibinfo {author} {\bibfnamefont {A.~D.}\ \bibnamefont {{Myers}}}, \bibinfo {author} {\bibfnamefont {P.}~\bibnamefont {{Noterdaeme}}}, \bibinfo {author} {\bibfnamefont {I.}~\bibnamefont {{P{\^a}ris}}}, \bibinfo {author} {\bibfnamefont {P.}~\bibnamefont {{Petitjean}}}, \bibinfo {author} {\bibfnamefont {M.~M.}\ \bibnamefont {{Pieri}}}, \bibinfo {author} {\bibfnamefont {J.}~\bibnamefont {{Rich}}}, \bibinfo {author}
  {\bibfnamefont {E.}~\bibnamefont {{Rollinde}}}, \bibinfo {author} {\bibfnamefont {N.~P.}\ \bibnamefont {{Ross}}}, \bibinfo {author} {\bibfnamefont {D.~J.}\ \bibnamefont {{Schlegel}}}, \bibinfo {author} {\bibfnamefont {D.~P.}\ \bibnamefont {{Schneider}}}, \bibinfo {author} {\bibfnamefont {A.}~\bibnamefont {{Slosar}}},\ and\ \bibinfo {author} {\bibfnamefont {D.~H.}\ \bibnamefont {{Weinberg}}},\ }\href {https://doi.org/10.1051/0004-6361/201322130} {\bibfield  {journal} {\bibinfo  {journal} {Astron. Astrophys.}\ }\textbf {\bibinfo {volume} {559}},\ \bibinfo {eid} {A85} (\bibinfo {year} {2013})},\ \Eprint {https://arxiv.org/abs/1306.5896} {arXiv:1306.5896 [astro-ph.CO]} \BibitemShut {NoStop}%
\bibitem [{\citenamefont {{Ir{\v{s}}i{\v{c}}}}\ \emph {et~al.}(2017{\natexlab{b}})\citenamefont {{Ir{\v{s}}i{\v{c}}}}, \citenamefont {{Viel}}, \citenamefont {{Berg}}, \citenamefont {{D'Odorico}}, \citenamefont {{Haehnelt}}, \citenamefont {{Cristiani}}, \citenamefont {{Cupani}}, \citenamefont {{Kim}}, \citenamefont {{L{\'o}pez}}, \citenamefont {{Ellison}}, \citenamefont {{Becker}}, \citenamefont {{Christensen}}, \citenamefont {{Denney}}, \citenamefont {{Worseck}},\ and\ \citenamefont {{Bolton}}}]{2017MNRAS.466.4332I}%
  \BibitemOpen
  \bibfield  {author} {\bibinfo {author} {\bibfnamefont {V.}~\bibnamefont {{Ir{\v{s}}i{\v{c}}}}}, \bibinfo {author} {\bibfnamefont {M.}~\bibnamefont {{Viel}}}, \bibinfo {author} {\bibfnamefont {T.~A.~M.}\ \bibnamefont {{Berg}}}, \bibinfo {author} {\bibfnamefont {V.}~\bibnamefont {{D'Odorico}}}, \bibinfo {author} {\bibfnamefont {M.~G.}\ \bibnamefont {{Haehnelt}}}, \bibinfo {author} {\bibfnamefont {S.}~\bibnamefont {{Cristiani}}}, \bibinfo {author} {\bibfnamefont {G.}~\bibnamefont {{Cupani}}}, \bibinfo {author} {\bibfnamefont {T.-S.}\ \bibnamefont {{Kim}}}, \bibinfo {author} {\bibfnamefont {S.}~\bibnamefont {{L{\'o}pez}}}, \bibinfo {author} {\bibfnamefont {S.}~\bibnamefont {{Ellison}}}, \bibinfo {author} {\bibfnamefont {G.~D.}\ \bibnamefont {{Becker}}}, \bibinfo {author} {\bibfnamefont {L.}~\bibnamefont {{Christensen}}}, \bibinfo {author} {\bibfnamefont {K.~D.}\ \bibnamefont {{Denney}}}, \bibinfo {author} {\bibfnamefont {G.}~\bibnamefont {{Worseck}}},\ and\ \bibinfo {author} {\bibfnamefont {J.~S.}\ \bibnamefont
  {{Bolton}}},\ }\href {https://doi.org/10.1093/mnras/stw3372} {\bibfield  {journal} {\bibinfo  {journal} {\mnras}\ }\textbf {\bibinfo {volume} {466}},\ \bibinfo {pages} {4332} (\bibinfo {year} {2017}{\natexlab{b}})},\ \Eprint {https://arxiv.org/abs/1702.01761} {arXiv:1702.01761 [astro-ph.CO]} \BibitemShut {NoStop}%
\bibitem [{\citenamefont {{Chabanier}}\ \emph {et~al.}(2019)\citenamefont {{Chabanier}}, \citenamefont {{Palanque-Delabrouille}}, \citenamefont {{Y{\`e}che}}, \citenamefont {{Le Goff}}, \citenamefont {{Armengaud}}, \citenamefont {{Bautista}}, \citenamefont {{Blomqvist}}, \citenamefont {{Busca}}, \citenamefont {{Dawson}}, \citenamefont {{Etourneau}}, \citenamefont {{Font-Ribera}}, \citenamefont {{Lee}}, \citenamefont {{du Mas des Bourboux}}, \citenamefont {{Pieri}}, \citenamefont {{Rich}}, \citenamefont {{Rossi}}, \citenamefont {{Schneider}},\ and\ \citenamefont {{Slosar}}}]{2019JCAP...07..017C}%
  \BibitemOpen
  \bibfield  {author} {\bibinfo {author} {\bibfnamefont {S.}~\bibnamefont {{Chabanier}}}, \bibinfo {author} {\bibfnamefont {N.}~\bibnamefont {{Palanque-Delabrouille}}}, \bibinfo {author} {\bibfnamefont {C.}~\bibnamefont {{Y{\`e}che}}}, \bibinfo {author} {\bibfnamefont {J.-M.}\ \bibnamefont {{Le Goff}}}, \bibinfo {author} {\bibfnamefont {E.}~\bibnamefont {{Armengaud}}}, \bibinfo {author} {\bibfnamefont {J.}~\bibnamefont {{Bautista}}}, \bibinfo {author} {\bibfnamefont {M.}~\bibnamefont {{Blomqvist}}}, \bibinfo {author} {\bibfnamefont {N.}~\bibnamefont {{Busca}}}, \bibinfo {author} {\bibfnamefont {K.}~\bibnamefont {{Dawson}}}, \bibinfo {author} {\bibfnamefont {T.}~\bibnamefont {{Etourneau}}}, \bibinfo {author} {\bibfnamefont {A.}~\bibnamefont {{Font-Ribera}}}, \bibinfo {author} {\bibfnamefont {Y.}~\bibnamefont {{Lee}}}, \bibinfo {author} {\bibfnamefont {H.}~\bibnamefont {{du Mas des Bourboux}}}, \bibinfo {author} {\bibfnamefont {M.}~\bibnamefont {{Pieri}}}, \bibinfo {author} {\bibfnamefont {J.}~\bibnamefont
  {{Rich}}}, \bibinfo {author} {\bibfnamefont {G.}~\bibnamefont {{Rossi}}}, \bibinfo {author} {\bibfnamefont {D.}~\bibnamefont {{Schneider}}},\ and\ \bibinfo {author} {\bibfnamefont {A.}~\bibnamefont {{Slosar}}},\ }\href {https://doi.org/10.1088/1475-7516/2019/07/017} {\bibfield  {journal} {\bibinfo  {journal} {JCAP}\ }\textbf {\bibinfo {volume} {2019}}\bibfield  {number} {\bibinfo  {number} { (7)},\ \bibinfo {eid} {017}},\ }\Eprint {https://arxiv.org/abs/1812.03554} {arXiv:1812.03554 [astro-ph.CO]} \BibitemShut {NoStop}%
\bibitem [{\citenamefont {{Kara{\c{c}}ayl{\i}}}\ \emph {et~al.}(2022)\citenamefont {{Kara{\c{c}}ayl{\i}}}, \citenamefont {{Padmanabhan}}, \citenamefont {{Font-Ribera}}, \citenamefont {{Ir{\v{s}}i{\v{c}}}}, \citenamefont {{Walther}}, \citenamefont {{Brooks}}, \citenamefont {{Gazta{\~n}aga}}, \citenamefont {{Kehoe}}, \citenamefont {{Levi}}, \citenamefont {{Ntelis}}, \citenamefont {{Palanque-Delabrouille}},\ and\ \citenamefont {{Tarl{\'e}}}}]{2022MNRAS.509.2842K}%
  \BibitemOpen
  \bibfield  {author} {\bibinfo {author} {\bibfnamefont {N.~G.}\ \bibnamefont {{Kara{\c{c}}ayl{\i}}}}, \bibinfo {author} {\bibfnamefont {N.}~\bibnamefont {{Padmanabhan}}}, \bibinfo {author} {\bibfnamefont {A.}~\bibnamefont {{Font-Ribera}}}, \bibinfo {author} {\bibfnamefont {V.}~\bibnamefont {{Ir{\v{s}}i{\v{c}}}}}, \bibinfo {author} {\bibfnamefont {M.}~\bibnamefont {{Walther}}}, \bibinfo {author} {\bibfnamefont {D.}~\bibnamefont {{Brooks}}}, \bibinfo {author} {\bibfnamefont {E.}~\bibnamefont {{Gazta{\~n}aga}}}, \bibinfo {author} {\bibfnamefont {R.}~\bibnamefont {{Kehoe}}}, \bibinfo {author} {\bibfnamefont {M.}~\bibnamefont {{Levi}}}, \bibinfo {author} {\bibfnamefont {P.}~\bibnamefont {{Ntelis}}}, \bibinfo {author} {\bibfnamefont {N.}~\bibnamefont {{Palanque-Delabrouille}}},\ and\ \bibinfo {author} {\bibfnamefont {G.}~\bibnamefont {{Tarl{\'e}}}},\ }\href {https://doi.org/10.1093/mnras/stab3201} {\bibfield  {journal} {\bibinfo  {journal} {\mnras}\ }\textbf {\bibinfo {volume} {509}},\ \bibinfo {pages} {2842}
  (\bibinfo {year} {2022})},\ \Eprint {https://arxiv.org/abs/2108.10870} {arXiv:2108.10870 [astro-ph.CO]} \BibitemShut {NoStop}%
\bibitem [{\citenamefont {{Day}}\ \emph {et~al.}(2019)\citenamefont {{Day}}, \citenamefont {{Tytler}},\ and\ \citenamefont {{Kambalur}}}]{2019MNRAS.489.2536D}%
  \BibitemOpen
  \bibfield  {author} {\bibinfo {author} {\bibfnamefont {A.}~\bibnamefont {{Day}}}, \bibinfo {author} {\bibfnamefont {D.}~\bibnamefont {{Tytler}}},\ and\ \bibinfo {author} {\bibfnamefont {B.}~\bibnamefont {{Kambalur}}},\ }\href {https://doi.org/10.1093/mnras/stz2214} {\bibfield  {journal} {\bibinfo  {journal} {\mnras}\ }\textbf {\bibinfo {volume} {489}},\ \bibinfo {pages} {2536} (\bibinfo {year} {2019})}\BibitemShut {NoStop}%
\bibitem [{\citenamefont {{DESI Collaboration}}\ and\ \citenamefont {{others}}(2016)}]{2016arXiv161100036D}%
  \BibitemOpen
  \bibfield  {author} {\bibinfo {author} {\bibnamefont {{DESI Collaboration}}}\ and\ \bibinfo {author} {\bibnamefont {{others}}},\ }\href {https://doi.org/10.48550/arXiv.1611.00036} {\bibfield  {journal} {\bibinfo  {journal} {arXiv e-prints}\ ,\ \bibinfo {eid} {arXiv:1611.00036}} (\bibinfo {year} {2016})},\ \Eprint {https://arxiv.org/abs/1611.00036} {arXiv:1611.00036 [astro-ph.IM]} \BibitemShut {NoStop}%
\bibitem [{\citenamefont {{DESI Collaboration}}\ and\ \citenamefont {{others}}(2023)}]{2023arXiv230606308D}%
  \BibitemOpen
  \bibfield  {author} {\bibinfo {author} {\bibnamefont {{DESI Collaboration}}}\ and\ \bibinfo {author} {\bibnamefont {{others}}},\ }\href {https://doi.org/10.48550/arXiv.2306.06308} {\bibfield  {journal} {\bibinfo  {journal} {arXiv e-prints}\ ,\ \bibinfo {eid} {arXiv:2306.06308}} (\bibinfo {year} {2023})},\ \Eprint {https://arxiv.org/abs/2306.06308} {arXiv:2306.06308 [astro-ph.CO]} \BibitemShut {NoStop}%
\bibitem [{\citenamefont {{Pieri}}\ \emph {et~al.}(2016)\citenamefont {{Pieri}}, \citenamefont {{Bonoli}}, \citenamefont {{Chaves-Montero}}, \citenamefont {{P{\^a}ris}}, \citenamefont {{Fumagalli}}, \citenamefont {{Bolton}}, \citenamefont {{Viel}}, \citenamefont {{Noterdaeme}}, \citenamefont {{Miralda-Escud{\'e}}}, \citenamefont {{Busca}}, \citenamefont {{Rahmani}}, \citenamefont {{Peroux}}, \citenamefont {{Font-Ribera}},\ and\ \citenamefont {{Trager}}}]{2016sf2a.conf..259P}%
  \BibitemOpen
  \bibfield  {author} {\bibinfo {author} {\bibfnamefont {M.~M.}\ \bibnamefont {{Pieri}}}, \bibinfo {author} {\bibfnamefont {S.}~\bibnamefont {{Bonoli}}}, \bibinfo {author} {\bibfnamefont {J.}~\bibnamefont {{Chaves-Montero}}}, \bibinfo {author} {\bibfnamefont {I.}~\bibnamefont {{P{\^a}ris}}}, \bibinfo {author} {\bibfnamefont {M.}~\bibnamefont {{Fumagalli}}}, \bibinfo {author} {\bibfnamefont {J.~S.}\ \bibnamefont {{Bolton}}}, \bibinfo {author} {\bibfnamefont {M.}~\bibnamefont {{Viel}}}, \bibinfo {author} {\bibfnamefont {P.}~\bibnamefont {{Noterdaeme}}}, \bibinfo {author} {\bibfnamefont {J.}~\bibnamefont {{Miralda-Escud{\'e}}}}, \bibinfo {author} {\bibfnamefont {N.~G.}\ \bibnamefont {{Busca}}}, \bibinfo {author} {\bibfnamefont {H.}~\bibnamefont {{Rahmani}}}, \bibinfo {author} {\bibfnamefont {C.}~\bibnamefont {{Peroux}}}, \bibinfo {author} {\bibfnamefont {A.}~\bibnamefont {{Font-Ribera}}},\ and\ \bibinfo {author} {\bibfnamefont {S.~C.}\ \bibnamefont {{Trager}}},\ }in\ \href
  {https://doi.org/10.48550/arXiv.1611.09388} {\emph {\bibinfo {booktitle} {SF2A-2016: Proceedings of the Annual meeting of the French Society of Astronomy and Astrophysics}}},\ \bibinfo {editor} {edited by\ \bibinfo {editor} {\bibfnamefont {C.}~\bibnamefont {{Reyl{\'e}}}}, \bibinfo {editor} {\bibfnamefont {J.}~\bibnamefont {{Richard}}}, \bibinfo {editor} {\bibfnamefont {L.}~\bibnamefont {{Cambr{\'e}sy}}}, \bibinfo {editor} {\bibfnamefont {M.}~\bibnamefont {{Deleuil}}}, \bibinfo {editor} {\bibfnamefont {E.}~\bibnamefont {{P{\'e}contal}}}, \bibinfo {editor} {\bibfnamefont {L.}~\bibnamefont {{Tresse}}},\ and\ \bibinfo {editor} {\bibfnamefont {I.}~\bibnamefont {{Vauglin}}}}\ (\bibinfo {year} {2016})\ pp.\ \bibinfo {pages} {259--266},\ \Eprint {https://arxiv.org/abs/1611.09388} {arXiv:1611.09388 [astro-ph.CO]} \BibitemShut {NoStop}%
\bibitem [{\citenamefont {{Bernardeau}}\ \emph {et~al.}(2002)\citenamefont {{Bernardeau}}, \citenamefont {{Mellier}},\ and\ \citenamefont {{van Waerbeke}}}]{2002A&A...389L..28B}%
  \BibitemOpen
  \bibfield  {author} {\bibinfo {author} {\bibfnamefont {F.}~\bibnamefont {{Bernardeau}}}, \bibinfo {author} {\bibfnamefont {Y.}~\bibnamefont {{Mellier}}},\ and\ \bibinfo {author} {\bibfnamefont {L.}~\bibnamefont {{van Waerbeke}}},\ }\href {https://doi.org/10.1051/0004-6361:20020700} {\bibfield  {journal} {\bibinfo  {journal} {\aap}\ }\textbf {\bibinfo {volume} {389}},\ \bibinfo {pages} {L28} (\bibinfo {year} {2002})},\ \Eprint {https://arxiv.org/abs/astro-ph/0201032} {arXiv:astro-ph/0201032 [astro-ph]} \BibitemShut {NoStop}%
\bibitem [{\citenamefont {{Takada}}\ and\ \citenamefont {{Jain}}(2003)}]{2003MNRAS.344..857T}%
  \BibitemOpen
  \bibfield  {author} {\bibinfo {author} {\bibfnamefont {M.}~\bibnamefont {{Takada}}}\ and\ \bibinfo {author} {\bibfnamefont {B.}~\bibnamefont {{Jain}}},\ }\href {https://doi.org/10.1046/j.1365-8711.2003.06868.x} {\bibfield  {journal} {\bibinfo  {journal} {\mnras}\ }\textbf {\bibinfo {volume} {344}},\ \bibinfo {pages} {857} (\bibinfo {year} {2003})},\ \Eprint {https://arxiv.org/abs/astro-ph/0304034} {arXiv:astro-ph/0304034 [astro-ph]} \BibitemShut {NoStop}%
\bibitem [{\citenamefont {{Semboloni}}\ \emph {et~al.}(2011)\citenamefont {{Semboloni}}, \citenamefont {{Schrabback}}, \citenamefont {{van Waerbeke}}, \citenamefont {{Vafaei}}, \citenamefont {{Hartlap}},\ and\ \citenamefont {{Hilbert}}}]{2011MNRAS.410..143S}%
  \BibitemOpen
  \bibfield  {author} {\bibinfo {author} {\bibfnamefont {E.}~\bibnamefont {{Semboloni}}}, \bibinfo {author} {\bibfnamefont {T.}~\bibnamefont {{Schrabback}}}, \bibinfo {author} {\bibfnamefont {L.}~\bibnamefont {{van Waerbeke}}}, \bibinfo {author} {\bibfnamefont {S.}~\bibnamefont {{Vafaei}}}, \bibinfo {author} {\bibfnamefont {J.}~\bibnamefont {{Hartlap}}},\ and\ \bibinfo {author} {\bibfnamefont {S.}~\bibnamefont {{Hilbert}}},\ }\href {https://doi.org/10.1111/j.1365-2966.2010.17430.x} {\bibfield  {journal} {\bibinfo  {journal} {\mnras}\ }\textbf {\bibinfo {volume} {410}},\ \bibinfo {pages} {143} (\bibinfo {year} {2011})},\ \Eprint {https://arxiv.org/abs/1005.4941} {arXiv:1005.4941 [astro-ph.CO]} \BibitemShut {NoStop}%
\bibitem [{\citenamefont {{Fu}}\ \emph {et~al.}(2014)\citenamefont {{Fu}}, \citenamefont {{Kilbinger}}, \citenamefont {{Erben}}, \citenamefont {{Heymans}}, \citenamefont {{Hildebrandt}}, \citenamefont {{Hoekstra}}, \citenamefont {{Kitching}}, \citenamefont {{Mellier}}, \citenamefont {{Miller}}, \citenamefont {{Semboloni}}, \citenamefont {{Simon}}, \citenamefont {{Van Waerbeke}}, \citenamefont {{Coupon}}, \citenamefont {{Harnois-D{\'e}raps}}, \citenamefont {{Hudson}}, \citenamefont {{Kuijken}}, \citenamefont {{Rowe}}, \citenamefont {{Schrabback}}, \citenamefont {{Vafaei}},\ and\ \citenamefont {{Velander}}}]{2014MNRAS.441.2725F}%
  \BibitemOpen
  \bibfield  {author} {\bibinfo {author} {\bibfnamefont {L.}~\bibnamefont {{Fu}}}, \bibinfo {author} {\bibfnamefont {M.}~\bibnamefont {{Kilbinger}}}, \bibinfo {author} {\bibfnamefont {T.}~\bibnamefont {{Erben}}}, \bibinfo {author} {\bibfnamefont {C.}~\bibnamefont {{Heymans}}}, \bibinfo {author} {\bibfnamefont {H.}~\bibnamefont {{Hildebrandt}}}, \bibinfo {author} {\bibfnamefont {H.}~\bibnamefont {{Hoekstra}}}, \bibinfo {author} {\bibfnamefont {T.~D.}\ \bibnamefont {{Kitching}}}, \bibinfo {author} {\bibfnamefont {Y.}~\bibnamefont {{Mellier}}}, \bibinfo {author} {\bibfnamefont {L.}~\bibnamefont {{Miller}}}, \bibinfo {author} {\bibfnamefont {E.}~\bibnamefont {{Semboloni}}}, \bibinfo {author} {\bibfnamefont {P.}~\bibnamefont {{Simon}}}, \bibinfo {author} {\bibfnamefont {L.}~\bibnamefont {{Van Waerbeke}}}, \bibinfo {author} {\bibfnamefont {J.}~\bibnamefont {{Coupon}}}, \bibinfo {author} {\bibfnamefont {J.}~\bibnamefont {{Harnois-D{\'e}raps}}}, \bibinfo {author} {\bibfnamefont {M.~J.}\ \bibnamefont {{Hudson}}},
  \bibinfo {author} {\bibfnamefont {K.}~\bibnamefont {{Kuijken}}}, \bibinfo {author} {\bibfnamefont {B.}~\bibnamefont {{Rowe}}}, \bibinfo {author} {\bibfnamefont {T.}~\bibnamefont {{Schrabback}}}, \bibinfo {author} {\bibfnamefont {S.}~\bibnamefont {{Vafaei}}},\ and\ \bibinfo {author} {\bibfnamefont {M.}~\bibnamefont {{Velander}}},\ }\href {https://doi.org/10.1093/mnras/stu754} {\bibfield  {journal} {\bibinfo  {journal} {\mnras}\ }\textbf {\bibinfo {volume} {441}},\ \bibinfo {pages} {2725} (\bibinfo {year} {2014})},\ \Eprint {https://arxiv.org/abs/1404.5469} {arXiv:1404.5469 [astro-ph.CO]} \BibitemShut {NoStop}%
\bibitem [{\citenamefont {{Ajani}}\ \emph {et~al.}(2023)\citenamefont {{Ajani}}, \citenamefont {{Harnois-D{\'e}raps}}, \citenamefont {{Pettorino}},\ and\ \citenamefont {{Starck}}}]{2023A&A...672L..10A}%
  \BibitemOpen
  \bibfield  {author} {\bibinfo {author} {\bibfnamefont {V.}~\bibnamefont {{Ajani}}}, \bibinfo {author} {\bibfnamefont {J.}~\bibnamefont {{Harnois-D{\'e}raps}}}, \bibinfo {author} {\bibfnamefont {V.}~\bibnamefont {{Pettorino}}},\ and\ \bibinfo {author} {\bibfnamefont {J.-L.}\ \bibnamefont {{Starck}}},\ }\href {https://doi.org/10.1051/0004-6361/202245510} {\bibfield  {journal} {\bibinfo  {journal} {\aap}\ }\textbf {\bibinfo {volume} {672}},\ \bibinfo {eid} {L10} (\bibinfo {year} {2023})},\ \Eprint {https://arxiv.org/abs/2211.10519} {arXiv:2211.10519 [astro-ph.CO]} \BibitemShut {NoStop}%
\bibitem [{\citenamefont {{Paillas}}\ \emph {et~al.}(2023)\citenamefont {{Paillas}}, \citenamefont {{Cuesta-Lazaro}}, \citenamefont {{Zarrouk}}, \citenamefont {{Cai}}, \citenamefont {{Percival}}, \citenamefont {{Nadathur}}, \citenamefont {{Pinon}}, \citenamefont {{de Mattia}},\ and\ \citenamefont {{Beutler}}}]{2023MNRAS.522..606P}%
  \BibitemOpen
  \bibfield  {author} {\bibinfo {author} {\bibfnamefont {E.}~\bibnamefont {{Paillas}}}, \bibinfo {author} {\bibfnamefont {C.}~\bibnamefont {{Cuesta-Lazaro}}}, \bibinfo {author} {\bibfnamefont {P.}~\bibnamefont {{Zarrouk}}}, \bibinfo {author} {\bibfnamefont {Y.-C.}\ \bibnamefont {{Cai}}}, \bibinfo {author} {\bibfnamefont {W.~J.}\ \bibnamefont {{Percival}}}, \bibinfo {author} {\bibfnamefont {S.}~\bibnamefont {{Nadathur}}}, \bibinfo {author} {\bibfnamefont {M.}~\bibnamefont {{Pinon}}}, \bibinfo {author} {\bibfnamefont {A.}~\bibnamefont {{de Mattia}}},\ and\ \bibinfo {author} {\bibfnamefont {F.}~\bibnamefont {{Beutler}}},\ }\href {https://doi.org/10.1093/mnras/stad1017} {\bibfield  {journal} {\bibinfo  {journal} {\mnras}\ }\textbf {\bibinfo {volume} {522}},\ \bibinfo {pages} {606} (\bibinfo {year} {2023})},\ \Eprint {https://arxiv.org/abs/2209.04310} {arXiv:2209.04310 [astro-ph.CO]} \BibitemShut {NoStop}%
\bibitem [{\citenamefont {Welling}(2005)}]{Welling2005RobustHO}%
  \BibitemOpen
  \bibfield  {author} {\bibinfo {author} {\bibfnamefont {M.}~\bibnamefont {Welling}},\ }in\ \href@noop {} {\emph {\bibinfo {booktitle} {International Conference on Artificial Intelligence and Statistics}}}\ (\bibinfo {year} {2005})\BibitemShut {NoStop}%
\bibitem [{\citenamefont {{Cheng}}(2021)}]{2021PhDT........26C}%
  \BibitemOpen
  \bibfield  {author} {\bibinfo {author} {\bibfnamefont {S.}~\bibnamefont {{Cheng}}},\ }\emph {\bibinfo {title} {{Astrophysics and cosmology with the scattering transform}}},\ \href@noop {} {Ph.D. thesis},\ \bibinfo  {school} {Johns Hopkins University, Maryland} (\bibinfo {year} {2021})\BibitemShut {NoStop}%
\bibitem [{\citenamefont {{Anden}}\ and\ \citenamefont {{Mallat}}(2014)}]{2014ITSP...62.4114A}%
  \BibitemOpen
  \bibfield  {author} {\bibinfo {author} {\bibfnamefont {J.}~\bibnamefont {{Anden}}}\ and\ \bibinfo {author} {\bibfnamefont {S.}~\bibnamefont {{Mallat}}},\ }\href {https://doi.org/10.1109/TSP.2014.2326991} {\bibfield  {journal} {\bibinfo  {journal} {IEEE Transactions on Signal Processing}\ }\textbf {\bibinfo {volume} {62}},\ \bibinfo {pages} {4114} (\bibinfo {year} {2014})},\ \Eprint {https://arxiv.org/abs/1304.6763} {arXiv:1304.6763 [cs.SD]} \BibitemShut {NoStop}%
\bibitem [{\citenamefont {And{\'e}n}\ and\ \citenamefont {Mallat}(2011)}]{Andn2011MultiscaleSF}%
  \BibitemOpen
  \bibfield  {author} {\bibinfo {author} {\bibfnamefont {J.}~\bibnamefont {And{\'e}n}}\ and\ \bibinfo {author} {\bibfnamefont {S.}~\bibnamefont {Mallat}},\ }in\ \href@noop {} {\emph {\bibinfo {booktitle} {International Society for Music Information Retrieval Conference}}}\ (\bibinfo {year} {2011})\BibitemShut {NoStop}%
\bibitem [{\citenamefont {Sifre}\ and\ \citenamefont {Mallat}(2013)}]{6619007}%
  \BibitemOpen
  \bibfield  {author} {\bibinfo {author} {\bibfnamefont {L.}~\bibnamefont {Sifre}}\ and\ \bibinfo {author} {\bibfnamefont {S.}~\bibnamefont {Mallat}},\ }in\ \href {https://doi.org/10.1109/CVPR.2013.163} {\emph {\bibinfo {booktitle} {2013 IEEE Conference on Computer Vision and Pattern Recognition}}}\ (\bibinfo {year} {2013})\ pp.\ \bibinfo {pages} {1233--1240}\BibitemShut {NoStop}%
\bibitem [{\citenamefont {{Mallat}}(2011)}]{2011arXiv1101.2286M}%
  \BibitemOpen
  \bibfield  {author} {\bibinfo {author} {\bibfnamefont {S.}~\bibnamefont {{Mallat}}},\ }\href {https://doi.org/10.48550/arXiv.1101.2286} {\bibfield  {journal} {\bibinfo  {journal} {arXiv e-prints}\ ,\ \bibinfo {eid} {arXiv:1101.2286}} (\bibinfo {year} {2011})},\ \Eprint {https://arxiv.org/abs/1101.2286} {arXiv:1101.2286 [math.FA]} \BibitemShut {NoStop}%
\bibitem [{\citenamefont {{Cheng}}\ \emph {et~al.}(2020)\citenamefont {{Cheng}}, \citenamefont {{Ting}}, \citenamefont {{M{\'e}nard}},\ and\ \citenamefont {{Bruna}}}]{2020MNRAS.499.5902C}%
  \BibitemOpen
  \bibfield  {author} {\bibinfo {author} {\bibfnamefont {S.}~\bibnamefont {{Cheng}}}, \bibinfo {author} {\bibfnamefont {Y.-S.}\ \bibnamefont {{Ting}}}, \bibinfo {author} {\bibfnamefont {B.}~\bibnamefont {{M{\'e}nard}}},\ and\ \bibinfo {author} {\bibfnamefont {J.}~\bibnamefont {{Bruna}}},\ }\href {https://doi.org/10.1093/mnras/staa3165} {\bibfield  {journal} {\bibinfo  {journal} {\mnras}\ }\textbf {\bibinfo {volume} {499}},\ \bibinfo {pages} {5902} (\bibinfo {year} {2020})},\ \Eprint {https://arxiv.org/abs/2006.08561} {arXiv:2006.08561 [astro-ph.CO]} \BibitemShut {NoStop}%
\bibitem [{\citenamefont {{Cheng }}\ and\ \citenamefont {{M{\'e}nard}}(2021)}]{2021MNRAS.507.1012C}%
  \BibitemOpen
  \bibfield  {author} {\bibinfo {author} {\bibfnamefont {S.}~\bibnamefont {{Cheng }}}\ and\ \bibinfo {author} {\bibfnamefont {B.}~\bibnamefont {{M{\'e}nard}}},\ }\href {https://doi.org/10.1093/mnras/stab2102} {\bibfield  {journal} {\bibinfo  {journal} {\mnras}\ }\textbf {\bibinfo {volume} {507}},\ \bibinfo {pages} {1012} (\bibinfo {year} {2021})},\ \Eprint {https://arxiv.org/abs/2103.09247} {arXiv:2103.09247 [astro-ph.CO]} \BibitemShut {NoStop}%
\bibitem [{\citenamefont {{Ribli}}\ \emph {et~al.}(2019)\citenamefont {{Ribli}}, \citenamefont {{Pataki}}, \citenamefont {{Zorrilla Matilla}}, \citenamefont {{Hsu}}, \citenamefont {{Haiman}},\ and\ \citenamefont {{Csabai}}}]{2019MNRAS.490.1843R}%
  \BibitemOpen
  \bibfield  {author} {\bibinfo {author} {\bibfnamefont {D.}~\bibnamefont {{Ribli}}}, \bibinfo {author} {\bibfnamefont {B.~{\'A}.}\ \bibnamefont {{Pataki}}}, \bibinfo {author} {\bibfnamefont {J.~M.}\ \bibnamefont {{Zorrilla Matilla}}}, \bibinfo {author} {\bibfnamefont {D.}~\bibnamefont {{Hsu}}}, \bibinfo {author} {\bibfnamefont {Z.}~\bibnamefont {{Haiman}}},\ and\ \bibinfo {author} {\bibfnamefont {I.}~\bibnamefont {{Csabai}}},\ }\href {https://doi.org/10.1093/mnras/stz2610} {\bibfield  {journal} {\bibinfo  {journal} {\mnras}\ }\textbf {\bibinfo {volume} {490}},\ \bibinfo {pages} {1843} (\bibinfo {year} {2019})},\ \Eprint {https://arxiv.org/abs/1902.03663} {arXiv:1902.03663 [astro-ph.CO]} \BibitemShut {NoStop}%
\bibitem [{\citenamefont {{Pedersen}}\ \emph {et~al.}(2022)\citenamefont {{Pedersen}}, \citenamefont {{Ho}},\ and\ \citenamefont {{Eickenberg}}}]{2022mla..confE..40P}%
  \BibitemOpen
  \bibfield  {author} {\bibinfo {author} {\bibfnamefont {C.}~\bibnamefont {{Pedersen}}}, \bibinfo {author} {\bibfnamefont {S.}~\bibnamefont {{Ho}}},\ and\ \bibinfo {author} {\bibfnamefont {M.}~\bibnamefont {{Eickenberg}}},\ }in\ \href {https://doi.org/10.48550/arXiv.2307.14362} {\emph {\bibinfo {booktitle} {Machine Learning for Astrophysics}}}\ (\bibinfo {year} {2022})\ p.~\bibinfo {pages} {40},\ \Eprint {https://arxiv.org/abs/2307.14362} {arXiv:2307.14362 [astro-ph.IM]} \BibitemShut {NoStop}%
\bibitem [{\citenamefont {{Bruna}}\ and\ \citenamefont {{Mallat}}(2012)}]{2012arXiv1203.1513B}%
  \BibitemOpen
  \bibfield  {author} {\bibinfo {author} {\bibfnamefont {J.}~\bibnamefont {{Bruna}}}\ and\ \bibinfo {author} {\bibfnamefont {S.}~\bibnamefont {{Mallat}}},\ }\href {https://doi.org/10.48550/arXiv.1203.1513} {\bibfield  {journal} {\bibinfo  {journal} {arXiv e-prints}\ ,\ \bibinfo {eid} {arXiv:1203.1513}} (\bibinfo {year} {2012})},\ \Eprint {https://arxiv.org/abs/1203.1513} {arXiv:1203.1513 [cs.CV]} \BibitemShut {NoStop}%
\bibitem [{\citenamefont {{Greig}}\ \emph {et~al.}(2022)\citenamefont {{Greig}}, \citenamefont {{Ting}},\ and\ \citenamefont {{Kaurov}}}]{2022MNRAS.513.1719G}%
  \BibitemOpen
  \bibfield  {author} {\bibinfo {author} {\bibfnamefont {B.}~\bibnamefont {{Greig}}}, \bibinfo {author} {\bibfnamefont {Y.-S.}\ \bibnamefont {{Ting}}},\ and\ \bibinfo {author} {\bibfnamefont {A.~A.}\ \bibnamefont {{Kaurov}}},\ }\href {https://doi.org/10.1093/mnras/stac977} {\bibfield  {journal} {\bibinfo  {journal} {\mnras}\ }\textbf {\bibinfo {volume} {513}},\ \bibinfo {pages} {1719} (\bibinfo {year} {2022})},\ \Eprint {https://arxiv.org/abs/2204.02544} {arXiv:2204.02544 [astro-ph.CO]} \BibitemShut {NoStop}%
\bibitem [{\citenamefont {{Valogiannis}}\ and\ \citenamefont {{Dvorkin}}(2022)}]{2022PhRvD.105j3534V}%
  \BibitemOpen
  \bibfield  {author} {\bibinfo {author} {\bibfnamefont {G.}~\bibnamefont {{Valogiannis}}}\ and\ \bibinfo {author} {\bibfnamefont {C.}~\bibnamefont {{Dvorkin}}},\ }\href {https://doi.org/10.1103/PhysRevD.105.103534} {\bibfield  {journal} {\bibinfo  {journal} {\prd}\ }\textbf {\bibinfo {volume} {105}},\ \bibinfo {eid} {103534} (\bibinfo {year} {2022})},\ \Eprint {https://arxiv.org/abs/2108.07821} {arXiv:2108.07821 [astro-ph.CO]} \BibitemShut {NoStop}%
\bibitem [{\citenamefont {{Lidz}}\ \emph {et~al.}(2010)\citenamefont {{Lidz}}, \citenamefont {{Faucher-Gigu{\`e}re}}, \citenamefont {{Dall'Aglio}}, \citenamefont {{McQuinn}}, \citenamefont {{Fechner}}, \citenamefont {{Zaldarriaga}}, \citenamefont {{Hernquist}},\ and\ \citenamefont {{Dutta}}}]{2010ApJ...718..199L}%
  \BibitemOpen
  \bibfield  {author} {\bibinfo {author} {\bibfnamefont {A.}~\bibnamefont {{Lidz}}}, \bibinfo {author} {\bibfnamefont {C.-A.}\ \bibnamefont {{Faucher-Gigu{\`e}re}}}, \bibinfo {author} {\bibfnamefont {A.}~\bibnamefont {{Dall'Aglio}}}, \bibinfo {author} {\bibfnamefont {M.}~\bibnamefont {{McQuinn}}}, \bibinfo {author} {\bibfnamefont {C.}~\bibnamefont {{Fechner}}}, \bibinfo {author} {\bibfnamefont {M.}~\bibnamefont {{Zaldarriaga}}}, \bibinfo {author} {\bibfnamefont {L.}~\bibnamefont {{Hernquist}}},\ and\ \bibinfo {author} {\bibfnamefont {S.}~\bibnamefont {{Dutta}}},\ }\href {https://doi.org/10.1088/0004-637X/718/1/199} {\bibfield  {journal} {\bibinfo  {journal} {\apj}\ }\textbf {\bibinfo {volume} {718}},\ \bibinfo {pages} {199} (\bibinfo {year} {2010})},\ \Eprint {https://arxiv.org/abs/0909.5210} {arXiv:0909.5210 [astro-ph.CO]} \BibitemShut {NoStop}%
\bibitem [{\citenamefont {{Gaikwad}}\ \emph {et~al.}(2021)\citenamefont {{Gaikwad}}, \citenamefont {{Srianand}}, \citenamefont {{Haehnelt}},\ and\ \citenamefont {{Choudhury}}}]{Gaikwad:2021}%
  \BibitemOpen
  \bibfield  {author} {\bibinfo {author} {\bibfnamefont {P.}~\bibnamefont {{Gaikwad}}}, \bibinfo {author} {\bibfnamefont {R.}~\bibnamefont {{Srianand}}}, \bibinfo {author} {\bibfnamefont {M.~G.}\ \bibnamefont {{Haehnelt}}},\ and\ \bibinfo {author} {\bibfnamefont {T.~R.}\ \bibnamefont {{Choudhury}}},\ }\href {https://doi.org/10.1093/mnras/stab2017} {\bibfield  {journal} {\bibinfo  {journal} {\mnras}\ }\textbf {\bibinfo {volume} {506}},\ \bibinfo {pages} {4389} (\bibinfo {year} {2021})},\ \Eprint {https://arxiv.org/abs/2009.00016} {arXiv:2009.00016 [astro-ph.CO]} \BibitemShut {NoStop}%
\bibitem [{\citenamefont {{Bird}}\ \emph {et~al.}(2019)\citenamefont {{Bird}}, \citenamefont {{Rogers}}, \citenamefont {{Peiris}}, \citenamefont {{Verde}}, \citenamefont {{Font-Ribera}},\ and\ \citenamefont {{Pontzen}}}]{2019JCAP...02..050B}%
  \BibitemOpen
  \bibfield  {author} {\bibinfo {author} {\bibfnamefont {S.}~\bibnamefont {{Bird}}}, \bibinfo {author} {\bibfnamefont {K.~K.}\ \bibnamefont {{Rogers}}}, \bibinfo {author} {\bibfnamefont {H.~V.}\ \bibnamefont {{Peiris}}}, \bibinfo {author} {\bibfnamefont {L.}~\bibnamefont {{Verde}}}, \bibinfo {author} {\bibfnamefont {A.}~\bibnamefont {{Font-Ribera}}},\ and\ \bibinfo {author} {\bibfnamefont {A.}~\bibnamefont {{Pontzen}}},\ }\href {https://doi.org/10.1088/1475-7516/2019/02/050} {\bibfield  {journal} {\bibinfo  {journal} {\jcap}\ }\textbf {\bibinfo {volume} {2019}},\ \bibinfo {eid} {050} (\bibinfo {year} {2019})},\ \Eprint {https://arxiv.org/abs/1812.04654} {arXiv:1812.04654 [astro-ph.CO]} \BibitemShut {NoStop}%
\bibitem [{\citenamefont {Feng}\ \emph {et~al.}(2018)\citenamefont {Feng}, \citenamefont {Bird}, \citenamefont {Anderson}, \citenamefont {Font-Ribera},\ and\ \citenamefont {Pedersen}}]{yu_feng_2018_1451799}%
  \BibitemOpen
  \bibfield  {author} {\bibinfo {author} {\bibfnamefont {Y.}~\bibnamefont {Feng}}, \bibinfo {author} {\bibfnamefont {S.}~\bibnamefont {Bird}}, \bibinfo {author} {\bibfnamefont {L.}~\bibnamefont {Anderson}}, \bibinfo {author} {\bibfnamefont {A.}~\bibnamefont {Font-Ribera}},\ and\ \bibinfo {author} {\bibfnamefont {C.}~\bibnamefont {Pedersen}},\ }\href {https://doi.org/10.5281/zenodo.1451799} {\bibinfo {title} {Mp-gadget/mp-gadget: A tag for getting a doi}} (\bibinfo {year} {2018})\BibitemShut {NoStop}%
\bibitem [{\citenamefont {{Bird}}(2017)}]{2017ascl.soft10012B}%
  \BibitemOpen
  \bibfield  {author} {\bibinfo {author} {\bibfnamefont {S.}~\bibnamefont {{Bird}}},\ }\href@noop {} {\bibinfo {title} {{FSFE: Fake Spectra Flux Extractor}}},\ \bibinfo {howpublished} {Astrophysics Source Code Library, record ascl:1710.012} (\bibinfo {year} {2017}),\ \Eprint {https://arxiv.org/abs/1710.012} {ascl:1710.012} \BibitemShut {NoStop}%
\bibitem [{\citenamefont {{Viel}}\ \emph {et~al.}(2004)\citenamefont {{Viel}}, \citenamefont {{Haehnelt}},\ and\ \citenamefont {{Springel}}}]{2004MNRAS.354..684V}%
  \BibitemOpen
  \bibfield  {author} {\bibinfo {author} {\bibfnamefont {M.}~\bibnamefont {{Viel}}}, \bibinfo {author} {\bibfnamefont {M.~G.}\ \bibnamefont {{Haehnelt}}},\ and\ \bibinfo {author} {\bibfnamefont {V.}~\bibnamefont {{Springel}}},\ }\href {https://doi.org/10.1111/j.1365-2966.2004.08224.x} {\bibfield  {journal} {\bibinfo  {journal} {\mnras}\ }\textbf {\bibinfo {volume} {354}},\ \bibinfo {pages} {684} (\bibinfo {year} {2004})},\ \Eprint {https://arxiv.org/abs/astro-ph/0404600} {arXiv:astro-ph/0404600 [astro-ph]} \BibitemShut {NoStop}%
\bibitem [{\citenamefont {{Kim}}\ \emph {et~al.}(2007)\citenamefont {{Kim}}, \citenamefont {{Bolton}}, \citenamefont {{Viel}}, \citenamefont {{Haehnelt}},\ and\ \citenamefont {{Carswell}}}]{2007MNRAS.382.1657K}%
  \BibitemOpen
  \bibfield  {author} {\bibinfo {author} {\bibfnamefont {T.~S.}\ \bibnamefont {{Kim}}}, \bibinfo {author} {\bibfnamefont {J.~S.}\ \bibnamefont {{Bolton}}}, \bibinfo {author} {\bibfnamefont {M.}~\bibnamefont {{Viel}}}, \bibinfo {author} {\bibfnamefont {M.~G.}\ \bibnamefont {{Haehnelt}}},\ and\ \bibinfo {author} {\bibfnamefont {R.~F.}\ \bibnamefont {{Carswell}}},\ }\href {https://doi.org/10.1111/j.1365-2966.2007.12406.x} {\bibfield  {journal} {\bibinfo  {journal} {\mnras}\ }\textbf {\bibinfo {volume} {382}},\ \bibinfo {pages} {1657} (\bibinfo {year} {2007})},\ \Eprint {https://arxiv.org/abs/0711.1862} {arXiv:0711.1862 [astro-ph]} \BibitemShut {NoStop}%
\bibitem [{\citenamefont {{Bruna}}\ \emph {et~al.}(2013)\citenamefont {{Bruna}}, \citenamefont {{Mallat}}, \citenamefont {{Bacry}},\ and\ \citenamefont {{Muzy}}}]{2013arXiv1311.4104B}%
  \BibitemOpen
  \bibfield  {author} {\bibinfo {author} {\bibfnamefont {J.}~\bibnamefont {{Bruna}}}, \bibinfo {author} {\bibfnamefont {S.}~\bibnamefont {{Mallat}}}, \bibinfo {author} {\bibfnamefont {E.}~\bibnamefont {{Bacry}}},\ and\ \bibinfo {author} {\bibfnamefont {J.-F.}\ \bibnamefont {{Muzy}}},\ }\href {https://doi.org/10.48550/arXiv.1311.4104} {\bibfield  {journal} {\bibinfo  {journal} {arXiv e-prints}\ ,\ \bibinfo {eid} {arXiv:1311.4104}} (\bibinfo {year} {2013})},\ \Eprint {https://arxiv.org/abs/1311.4104} {arXiv:1311.4104 [stat.ME]} \BibitemShut {NoStop}%
\bibitem [{\citenamefont {{Andreux}}\ \emph {et~al.}(2018)\citenamefont {{Andreux}}, \citenamefont {{Angles}}, \citenamefont {{Exarchakis}}, \citenamefont {{Leonarduzzi}}, \citenamefont {{Rochette}}, \citenamefont {{Thiry}}, \citenamefont {{Zarka}}, \citenamefont {{Mallat}}, \citenamefont {{and{\'e}n}}, \citenamefont {{Belilovsky}}, \citenamefont {{Bruna}}, \citenamefont {{Lostanlen}}, \citenamefont {{Chaudhary}}, \citenamefont {{Hirn}}, \citenamefont {{Oyallon}}, \citenamefont {{Zhang}}, \citenamefont {{Cella}},\ and\ \citenamefont {{Eickenberg}}}]{2018arXiv181211214A}%
  \BibitemOpen
  \bibfield  {author} {\bibinfo {author} {\bibfnamefont {M.}~\bibnamefont {{Andreux}}}, \bibinfo {author} {\bibfnamefont {T.}~\bibnamefont {{Angles}}}, \bibinfo {author} {\bibfnamefont {G.}~\bibnamefont {{Exarchakis}}}, \bibinfo {author} {\bibfnamefont {R.}~\bibnamefont {{Leonarduzzi}}}, \bibinfo {author} {\bibfnamefont {G.}~\bibnamefont {{Rochette}}}, \bibinfo {author} {\bibfnamefont {L.}~\bibnamefont {{Thiry}}}, \bibinfo {author} {\bibfnamefont {J.}~\bibnamefont {{Zarka}}}, \bibinfo {author} {\bibfnamefont {S.}~\bibnamefont {{Mallat}}}, \bibinfo {author} {\bibfnamefont {J.}~\bibnamefont {{and{\'e}n}}}, \bibinfo {author} {\bibfnamefont {E.}~\bibnamefont {{Belilovsky}}}, \bibinfo {author} {\bibfnamefont {J.}~\bibnamefont {{Bruna}}}, \bibinfo {author} {\bibfnamefont {V.}~\bibnamefont {{Lostanlen}}}, \bibinfo {author} {\bibfnamefont {M.}~\bibnamefont {{Chaudhary}}}, \bibinfo {author} {\bibfnamefont {M.~J.}\ \bibnamefont {{Hirn}}}, \bibinfo {author} {\bibfnamefont {E.}~\bibnamefont {{Oyallon}}}, \bibinfo
  {author} {\bibfnamefont {S.}~\bibnamefont {{Zhang}}}, \bibinfo {author} {\bibfnamefont {C.}~\bibnamefont {{Cella}}},\ and\ \bibinfo {author} {\bibfnamefont {M.}~\bibnamefont {{Eickenberg}}},\ }\href {https://doi.org/10.48550/arXiv.1812.11214} {\bibfield  {journal} {\bibinfo  {journal} {arXiv e-prints}\ ,\ \bibinfo {eid} {arXiv:1812.11214}} (\bibinfo {year} {2018})},\ \Eprint {https://arxiv.org/abs/1812.11214} {arXiv:1812.11214 [cs.LG]} \BibitemShut {NoStop}%
\bibitem [{\citenamefont {{Lee}}\ \emph {et~al.}(2012)\citenamefont {{Lee}}, \citenamefont {{Suzuki}},\ and\ \citenamefont {{Spergel}}}]{Lee12}%
  \BibitemOpen
  \bibfield  {author} {\bibinfo {author} {\bibfnamefont {K.-G.}\ \bibnamefont {{Lee}}}, \bibinfo {author} {\bibfnamefont {N.}~\bibnamefont {{Suzuki}}},\ and\ \bibinfo {author} {\bibfnamefont {D.~N.}\ \bibnamefont {{Spergel}}},\ }\href {https://doi.org/10.1088/0004-6256/143/2/51} {\bibfield  {journal} {\bibinfo  {journal} {\aj}\ }\textbf {\bibinfo {volume} {143}},\ \bibinfo {eid} {51} (\bibinfo {year} {2012})},\ \Eprint {https://arxiv.org/abs/1108.6080} {arXiv:1108.6080 [astro-ph.CO]} \BibitemShut {NoStop}%
\bibitem [{\citenamefont {{Qezlou}}\ \emph {et~al.}(2022)\citenamefont {{Qezlou}}, \citenamefont {{Newman}}, \citenamefont {{Rudie}},\ and\ \citenamefont {{Bird}}}]{Qezlou22}%
  \BibitemOpen
  \bibfield  {author} {\bibinfo {author} {\bibfnamefont {M.}~\bibnamefont {{Qezlou}}}, \bibinfo {author} {\bibfnamefont {A.~B.}\ \bibnamefont {{Newman}}}, \bibinfo {author} {\bibfnamefont {G.~C.}\ \bibnamefont {{Rudie}}},\ and\ \bibinfo {author} {\bibfnamefont {S.}~\bibnamefont {{Bird}}},\ }\href {https://doi.org/10.3847/1538-4357/ac6259} {\bibfield  {journal} {\bibinfo  {journal} {\apj}\ }\textbf {\bibinfo {volume} {930}},\ \bibinfo {eid} {109} (\bibinfo {year} {2022})},\ \Eprint {https://arxiv.org/abs/2112.03930} {arXiv:2112.03930 [astro-ph.GA]} \BibitemShut {NoStop}%
\bibitem [{\citenamefont {{Lyke}}\ \emph {et~al.}(2020)\citenamefont {{Lyke}}, \citenamefont {{Higley}}, \citenamefont {{McLane}}, \citenamefont {{Schurhammer}}, \citenamefont {{Myers}}, \citenamefont {{Ross}}, \citenamefont {{Dawson}}, \citenamefont {{Chabanier}}, \citenamefont {{Martini}}, \citenamefont {{Busca}}, \citenamefont {{Mas des Bourboux}}, \citenamefont {{Salvato}}, \citenamefont {{Streblyanska}}, \citenamefont {{Zarrouk}}, \citenamefont {{Burtin}}, \citenamefont {{Anderson}}, \citenamefont {{Bautista}}, \citenamefont {{Bizyaev}}, \citenamefont {{Brandt}}, \citenamefont {{Brinkmann}}, \citenamefont {{Brownstein}}, \citenamefont {{Comparat}}, \citenamefont {{Green}}, \citenamefont {{de la Macorra}}, \citenamefont {{Mu{\~n}oz Guti{\'e}rrez}}, \citenamefont {{Hou}}, \citenamefont {{Newman}}, \citenamefont {{Palanque-Delabrouille}}, \citenamefont {{P{\^a}ris}}, \citenamefont {{Percival}}, \citenamefont {{Petitjean}}, \citenamefont {{Rich}}, \citenamefont {{Rossi}}, \citenamefont {{Schneider}},
  \citenamefont {{Smith}}, \citenamefont {{Vivek}},\ and\ \citenamefont {{Weaver}}}]{SDSS-DR16Q}%
  \BibitemOpen
  \bibfield  {author} {\bibinfo {author} {\bibfnamefont {B.~W.}\ \bibnamefont {{Lyke}}}, \bibinfo {author} {\bibfnamefont {A.~N.}\ \bibnamefont {{Higley}}}, \bibinfo {author} {\bibfnamefont {J.~N.}\ \bibnamefont {{McLane}}}, \bibinfo {author} {\bibfnamefont {D.~P.}\ \bibnamefont {{Schurhammer}}}, \bibinfo {author} {\bibfnamefont {A.~D.}\ \bibnamefont {{Myers}}}, \bibinfo {author} {\bibfnamefont {A.~J.}\ \bibnamefont {{Ross}}}, \bibinfo {author} {\bibfnamefont {K.}~\bibnamefont {{Dawson}}}, \bibinfo {author} {\bibfnamefont {S.}~\bibnamefont {{Chabanier}}}, \bibinfo {author} {\bibfnamefont {P.}~\bibnamefont {{Martini}}}, \bibinfo {author} {\bibfnamefont {N.~G.}\ \bibnamefont {{Busca}}}, \bibinfo {author} {\bibfnamefont {H.~d.}\ \bibnamefont {{Mas des Bourboux}}}, \bibinfo {author} {\bibfnamefont {M.}~\bibnamefont {{Salvato}}}, \bibinfo {author} {\bibfnamefont {A.}~\bibnamefont {{Streblyanska}}}, \bibinfo {author} {\bibfnamefont {P.}~\bibnamefont {{Zarrouk}}}, \bibinfo {author} {\bibfnamefont {E.}~\bibnamefont
  {{Burtin}}}, \bibinfo {author} {\bibfnamefont {S.~F.}\ \bibnamefont {{Anderson}}}, \bibinfo {author} {\bibfnamefont {J.}~\bibnamefont {{Bautista}}}, \bibinfo {author} {\bibfnamefont {D.}~\bibnamefont {{Bizyaev}}}, \bibinfo {author} {\bibfnamefont {W.~N.}\ \bibnamefont {{Brandt}}}, \bibinfo {author} {\bibfnamefont {J.}~\bibnamefont {{Brinkmann}}}, \bibinfo {author} {\bibfnamefont {J.~R.}\ \bibnamefont {{Brownstein}}}, \bibinfo {author} {\bibfnamefont {J.}~\bibnamefont {{Comparat}}}, \bibinfo {author} {\bibfnamefont {P.}~\bibnamefont {{Green}}}, \bibinfo {author} {\bibfnamefont {A.}~\bibnamefont {{de la Macorra}}}, \bibinfo {author} {\bibfnamefont {A.}~\bibnamefont {{Mu{\~n}oz Guti{\'e}rrez}}}, \bibinfo {author} {\bibfnamefont {J.}~\bibnamefont {{Hou}}}, \bibinfo {author} {\bibfnamefont {J.~A.}\ \bibnamefont {{Newman}}}, \bibinfo {author} {\bibfnamefont {N.}~\bibnamefont {{Palanque-Delabrouille}}}, \bibinfo {author} {\bibfnamefont {I.}~\bibnamefont {{P{\^a}ris}}}, \bibinfo {author} {\bibfnamefont {W.~J.}\
  \bibnamefont {{Percival}}}, \bibinfo {author} {\bibfnamefont {P.}~\bibnamefont {{Petitjean}}}, \bibinfo {author} {\bibfnamefont {J.}~\bibnamefont {{Rich}}}, \bibinfo {author} {\bibfnamefont {G.}~\bibnamefont {{Rossi}}}, \bibinfo {author} {\bibfnamefont {D.~P.}\ \bibnamefont {{Schneider}}}, \bibinfo {author} {\bibfnamefont {A.}~\bibnamefont {{Smith}}}, \bibinfo {author} {\bibfnamefont {M.}~\bibnamefont {{Vivek}}},\ and\ \bibinfo {author} {\bibfnamefont {B.~A.}\ \bibnamefont {{Weaver}}},\ }\href {https://doi.org/10.3847/1538-4365/aba623} {\bibfield  {journal} {\bibinfo  {journal} {\apjs}\ }\textbf {\bibinfo {volume} {250}},\ \bibinfo {eid} {8} (\bibinfo {year} {2020})},\ \Eprint {https://arxiv.org/abs/2007.09001} {arXiv:2007.09001 [astro-ph.GA]} \BibitemShut {NoStop}%
\bibitem [{\citenamefont {{Fernandez}}\ \emph {et~al.}(2023)\citenamefont {{Fernandez}}, \citenamefont {{Bird}},\ and\ \citenamefont {{Ho}}}]{2023arXiv230903943F}%
  \BibitemOpen
  \bibfield  {author} {\bibinfo {author} {\bibfnamefont {M.~A.}\ \bibnamefont {{Fernandez}}}, \bibinfo {author} {\bibfnamefont {S.}~\bibnamefont {{Bird}}},\ and\ \bibinfo {author} {\bibfnamefont {M.-F.}\ \bibnamefont {{Ho}}},\ }\href {https://doi.org/10.48550/arXiv.2309.03943} {\bibfield  {journal} {\bibinfo  {journal} {arXiv e-prints}\ ,\ \bibinfo {eid} {arXiv:2309.03943}} (\bibinfo {year} {2023})},\ \Eprint {https://arxiv.org/abs/2309.03943} {arXiv:2309.03943 [astro-ph.CO]} \BibitemShut {NoStop}%
\end{thebibliography}%

%\section*{Appendix}
%\subsection*{1d Wavelet Scatter Transform}
%\begin{figure*}[ht]
%\includegraphics[width=\linewidth]{zeroth_1d.pdf} 
%\includegraphics[width=\linewidth]{first_1d.pdf} 
%\includegraphics[width=\linewidth]{sec_1d.pdf} 
%\caption{WST applied to fiducial model for Lyman-$\alpha$ forest to generate scattering coefficients}
%\label{gen_wst}
%\end{figure*}

\end{document}